# Bilayer Low-Density Parity-Check Codes for Decode-and-Forward in Relay Channels

Peyman Razaghi, *Student Member, IEEE*, and Wei Yu, *Member, IEEE*

## Abstract

This paper describes an efficient implementation of binning for the relay channel using low-density parity-check (LDPC) codes. We devise bilayer LDPC codes to approach the theoretically promised rate of the decode-and-forward relaying strategy by incorporating relay-generated information bits in specially designed bilayer graphical code structures. While conventional LDPC codes are sensitively tuned to operate efficiently at a certain channel parameter, the proposed bilayer LDPC codes are capable of working at two different channel parameters and two different rates: that at the relay and at the destination. To analyze the performance of bilayer LDPC codes, bilayer density evolution is devised as an extension of the standard density evolution algorithm. Based on bilayer density evolution, a design methodology is developed for the bilayer codes in which the degree distribution is iteratively improved using linear programming. Further, in order to approach the theoretical decode-and-forward rate for a wide range of channel parameters, this paper proposes two different forms bilayer codes, the bilayer-expurgated and bilayer-lengthened codes. It is demonstrated that a properly designed bilayer LDPC code can achieve an asymptotic infinite-length threshold within 0.24 dB gap to the Shannon limits of two different channels simultaneously for a wide range of channel parameters. By practical code construction, finite-length bilayer codes are shown to be able to approach within a 0.6 dB gap to the theoretical decode-and-forward rate of the relay channel at a block length of $10^5$ and a bit-error probability (BER) of $10^{-4}$. Finally, it is demonstrated that a generalized version of the proposed bilayer code construction is applicable to relay networks with multiple relays.

Manuscript submitted to the *IEEE Transactions on Information Theory* on September 1, 2006. The material in this paper has been presented in part at the *IEEE International Conference on Communications (ICC)*, in Istanbul, Turkey, June 2006. The authors are with The Edward S. Rogers Sr. Department of Electrical and Computer Engineering, University of Toronto, 10 King's College Road, Toronto, Ontario M5S 3G4, Canada. e-mails: peyman@comm.utoronto.ca, weiyu@comm.utoronto.ca. Phone: 416-946-8665. FAX: 416-978-4425. This work was supported by the Natural Science and Engineering Research Council (NSERC) of Canada under a discovery grant and under the Canada Research Chairs program. Kindly address correspondence to Peyman Razaghi (peyman@comm.utoronto.ca).

# I. INTRODUCTION

Low-density parity-check (LDPC) codes have proved to be very powerful in approaching the capacity of conventional single-user communication channels. The key idea of LDPC codes is to practically implement the random coding theorem of Shannon by enforcing a set of random parity-check constraints on information bits. While random coding is a fundamental element of single-user information theory, binning is of fundamental importance in multiuser scenarios. In this paper, we explore the possibility of using LDPC codes to practically implement binning and to approach the theoretical results derived by random binning and random coding arguments for an important example of multi-user channels: the relay channel.

In a relay channel, a single source $X$ attempts to communicate to a single destination $Y$ with the help of a relay. The relay receives $Y_1$ and sends out $X_1$ based on $Y_1$. The relay channel is defined by the joint distribution $p(y, y_1|x, x_1)$. A schematic of the relay channel is illustrated in Fig. 1. Although the capacity of the relay channel is still an open problem, several clever methods have been designed to take advantage of the information available at the relay. The classic work of Cover and El Gamal [1] describes two basic strategies: first, a decode-and-forward strategy in which the relay completely decodes the transmitted message and partially forwards the decoded message using a binning technique to allow the complete resolution of the message at the decoder, and second, a more complex quantize-and-forward strategy in which the relay does not need to decode the source's message. Both strategies rely on a block-Markov coding scheme in which each coding block consists of simultaneous decoding (or quantizing) of the current block at the relay and the decoding of the previous block at the destination. Cover and El Gamal [1] proved that the decode-and-forward strategy is capacity achieving for a class of degraded relay channels.

This paper focuses on practical implementation of the decode-and-forward strategy for the relay channel. We restrict our attention to Gaussian relay channels at low signal-to-noise ratios (SNRs) for which binary linear codes are suitable. We show that, within a linear coding framework, the binning strategy in which a bin index of the codeword is transmitted by the relay to the destination can be interpreted as a *parity-forwarding* scheme. Further, the optimal code design for the decode-and-forward strategy entails to the design of a LDPC code working at two *different* channel SNRs: a high SNR at the relay and a low SNR at the destination. This represents novel LDPC code constructions, named *bilayer LDPC codes* in this paper.

The main results of our work are as follows. We propose two new ensembles of LDPC codes, bilayer-expurgated codes and bilayer-lengthened LDPC codes, both with an embedding structure, to simultaneously approach the capacities of two Gaussian channels at two different SNRs. The performance analysis and design methodologies for these new ensembles of bilayer LDPC codes are developed by generalizing density evolution [2] for standard LDPC codes to bilayer codes. We develop a design technique based on linear programming to optimize the variable degrees of the bilayer code. The two forms of bilayer code structure are necessary in order to accommodate the optimization of check degrees. Together, we show that specially structured irregular bilayer LDPC code with carefully chosen variable and check degree sequences can approach the theoretical decode-and-forward rate of the relay channel to within a fraction of dB for a wide range of different channel conditions. Finally, we generalize our code design for relay networks with multiple relays and show that our approach is applicable to general networks.

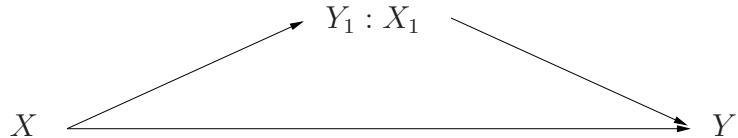

Fig. 1. The relay channel

The names bilayer-expurgated and bilayer-lengthened codes come from the linear coding terminology where *expurgating* refers to reducing the codebook size by increasing the number of parity-check equations while keeping the codeword length fixed, and *lengthening* refers to increasing the codebook size by increasing the codeword length while keeping the number of parity-check equations same [3]. In expurgating, codewords that do not satisfy the extra check equations are deleted from the codebook. The decode-and-forward operation in a relay channel can be thought of as an expurgating process in which the relay decodes the source codeword, then expurgates the source codebook by transmitting the extra parity bits (corresponding to the decoded codeword) to the destination. The code design problem for decode-and-forward then amounts to designing a code that is capacity-approaching at two different rates, both before and after expurgation.

Alternatively, the decode-and-forward operation can also be implemented using the code lengthening process, in which the lengthened code is decoded at the relay, and the extra variable bits are transmitted to the destination. The destination then decodes a *shortened* code at a lower rate. Again, the code design problem amounts to designing a code which is capacity-approaching at two different rates, before and after the lengthening process.

This paper focuses on novel design methodologies for bilayer-expurgated and bilayer-lengthened codes. As illustrated later, depending on the relative SNRs at the relay and at the destination, the expurgated code may be more appropriate than the lengthened code, or vice versa. Together, these two techniques are capable of covering a wide range of channel SNRs.

## A. Related Work

Recent interests in wireless ad-hoc and sensor networks have fueled a new surge of research activities both on the capacity (e.g. [4], [5], [6], [7]) and on practical implementation of communication schemes (e.g. [8], [9], [10]) for the relay channel in the wireless setting. Earlier work on coding for the relay channel includes the implementation of the decode-and-forward strategy using LDPC codes [11] and turbo codes [12], where performances approaching 1-1.5dB of the theoretical limit have been reported. The work of [11] employs the so-called regular encoding method for the relay channel. In this case, the source encodes its message using a LDPC code; the relay decodes the source message and retransmits the entire source codeword using a second LDPC code; the destination decodes the source message jointly over the combined graph. Our approach is different in that an irregular coding method based on the original binning technique of Cover and El Gamal is implemented. In this case, the relay forwards the bin index of the source codeword to the destination. Although both regular and irregular coding can achieve the full decode-and-forward rate, irregular coding is more flexible. For example, it can be applied when the relay-destination link is a digital link; it is also more easily generalizable to relay networks with multiple relays, as shown later in Section VII; see also [13]. The LDPC code design problem for regular coding has recently been considered in [14]. The present work deals with the code design problem for irregular coding.

This work is also related to a large number of recent and independent work on the application of LDPC codes to full duplex and half duplex relay channels [15], [16], [17]. In [15], an irregular expurgated coding protocol is devised, and a density evolution approach is used for code design. The code analysis and design methodology of [15] is based on conventional density evolution, where the performance of bilayer codes are approximated by that of the ensemble of standard LDPC codes. Using such an approximation, [15] reports a gap of 1dB to capacity. The work of [16] considers the use of independent source and relay codebooks in a relay channel, which is applicable only when the relay has excess power. In this case, the successful decoding of the expurgated code at the destination is easy to guarantee, and the optimal design of the bilayer-expurgated code simplifies to that of a conventional LDPC code. The work [17] applies conventional LDPC codes to the half-duplex relay channel and proposes a *random* puncturing scheme for LDPC codes optimized for single-user channels.

The present work differs from [15], [16], [17] in that bilayer LDPC codes capable of approaching capacities at two different rates are explicitly designed by expurgating and lengthening conventional LDPC codes in a structured manner. We show that in order to truly approach the decode-and-forward rate for the relay channel, *new ensembles* of LDPC codes with a carefully designed bilayer structure should be used. For bilayer-expurgated LDPC codes, the random code ensemble is parameterized by doubly-indexed degree distribution sequences on the variables; for bilayer-lengthened LDPC codes the check degree distribution has double indices. While the use of conventional LDPC codes may approach the theoretical limit to about 1-1.5dB, our specific design is asymptotically capable of closing the gap to 0.24dB for a wide range of channel parameters.

This work differs from the code design methodologies of [14], [15], [17] in that we devise an iterative degree-distribution update method based on linear programming for the bilayer codes. The iterative method optimizes the decoder output error probability based on the result of exact density evolution in every step. This approach is based on the design technique in [18] and is inspired by LDPC code design methods based on extrinsic information transfer (EXIT) charts [19], [20], [21], [22], [23], but is more accurate, since no Gaussian approximation is used in density evolution. By practical code construction, we show that the methodology is capable of approaching the decode-and-forward rates for a variety of channel conditions.

The code design problem for the relay channel is related to the general concept of *rateless* codes, popularized by the invention of fountain codes [24] and, more recently, raptor codes [25], which are capable of approaching the capacities of binary erasure channels (BEC) irrespective of the erasure probability. Bilayer LDPC codes for the relay channel is similar to rateless codes in that both are capable of working at multiple different rates. However, the code design requirement for the relay channel is also fundamentally different in the following aspect. In a relay channel, the extra information for the second code is transmitted via a separately coded channel from the relay to the destination. In contrast, the channel model for rateless codes typically assumes that additional bits are sent through the same channel (thus are corrupted by the channel noise.) In addition, fountain codes and raptor codes are designed specifically for the BEC; whether practical rateless codes exist for more general channel models is still an open research issue [26]. In this sense, the design methodology for fountain codes and raptor codes cannot be directly applied to the general relay channel, (except in the erasure case, which is elaborated later in the paper.)

Finally, the code construction proposed in this paper is also related to the use of punctured rate-compatible LDPC codes for incremental redundancy hybrid automatic repeat request (IR-HARQ) protocols for wireless transmission channels (e.g. [27], [28], [29], [30], [31].) In the HARQ setting, additional coded data bits are sent when the decoder fails to decode. Again, the additional bits can be potentially corrupted by the same channel, resulting in a different coding problem as compared to the relay setting. Thus, the existing HARQ coding schemes are not directly applicable for decode-and-forward.

*B. Outline of the Paper*

The rest of this paper is organized as follows. Section II provides a brief review of Cover and El Gamal's decode-and-forward scheme, then proposes the bilayer LDPC code structure based on expurgating. After a review of density evolution and LDPC code optimization methods for conventional LDPC codes in Section III, bilayer density evolution and the proposed linear-programming-based design methodology for bilayer-expurgated LDPC codes are developed in Section IV. Section V proposes the bilayer-lengthened LDPC codes and describes the associated analysis and design tools. Code construction and numerical results are given in Section VI for a range of relay channel parameters. Section VII provides a generalization of bilayer codes to multiple-relay networks. Finally, concluding remarks are made in Section VIII.

## II. Coding for Decode-and-Forward

*A. Decode-and-Forward Strategy*

This section briefly reviews the decode-and-forward strategy of [1, Theorem 1]. In the block-Markov decode-and-forward scheme, transmissions occur in successive blocks. In each block $i$, the source and the relay send two messages to the destination: the source's data message denoted by $w_i \in \{1, 2, \ldots, 2^{nR}\}$ (which is encoded through the random variable $X$) and the relay's message $s_i \in \{1, 2, \cdots, 2^{nR_1}\}$ (which is encoded through the random variable $X_1$.) The source rate, $R$, is such that the relay is able to decode $w_i$ with an arbitrarily low error probability; however, the destination is unable to uniquely decode $w_i$ because of its poorer channel. The relay's message, $s_i$, helps the destination decode $w_{i-1}$ in block $i$ by restricting $w_{i-1}$ to be inside a bin of size $2^{n(R-R_1)}$. Let $\mathcal{B} = \{\mathcal{S}_1, \mathcal{S}_2, \cdots, \mathcal{S}_{2^{nR_1}}\}$ be a random uniform partition of the set $\{1, 2, \cdots, 2^{nR}\}$ into $2^{nR_1}$ bins of size $2^{n(R-R_1)}$. The relay's message, $s_i$, is determined as the index of the bin in which $w_{i-1}$ falls, i.e., $w_{i-1} \in \mathcal{S}_{s_i}$.

Random codebooks to transmit $s$ and $w$ are constructed as follows. Assume that in block $i$, both the source and the relay know $s_i$; this is a valid assumption, since $s_i$ is determined by $w_{i-1}$. The source uses different codebooks for each different $s_i$. To encode $w_i$, the source utilizes a random codebook $\mathcal{X}(\cdot|s_i)$ of size $2^{nR}$ generated according to the probability distribution $p(x|x_1)$ and transmits the codeword $X(w_i|s_i)$.

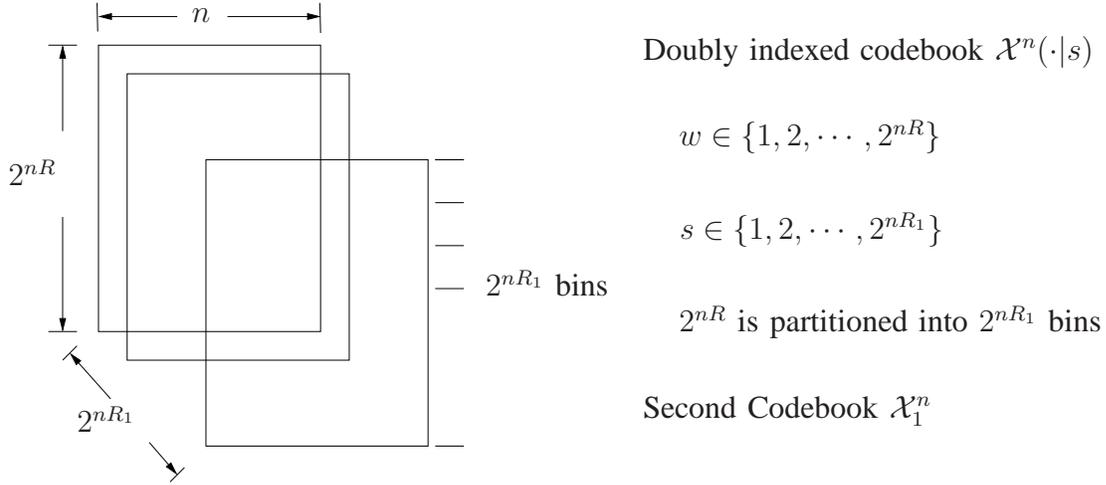

Doubly indexed codebook $\mathcal{X}^n(\cdot|s)$

$w \in \{1, 2, \cdots, 2^{nR}\}$

$s \in \{1, 2, \cdots, 2^{nR_1}\}$

$2^{nR}$ is partitioned into $2^{nR_1}$ bins

Second Codebook $\mathcal{X}_1^n$

Fig. 2. Codebook construction for the relay channel

In block $i$, the relay sends $s_i$ by transmitting the codeword $X_1(s_i)$ of the random codebook $\mathcal{X}_1$ of size $2^{nR_1}$ generated according to the probability distribution $p(x_1)$. Thus, while the relay uses an independent codebook to encode $s_i$, the source codebook is *doubly indexed*, by both $w_i$ and $s_i$, in order to facilitate cooperative transmission of $s_i$ to the destination. A key feature of the decode-and-forward schemes is that the source codebook must be decodable both at the relay and at the destination. In block $i$, the relay decodes $w_i$ and computes $s_i$. The destination decodes $w_{i-1}$ with the help of $s_i$. This double decodability condition gives rise to a bilayer code structure.

The decode-and-forward rate for the relay channel is computed as follows. In block $i$, the relay decodes $w_i$ which is possible if

$$R < I(X; Y_1 | X_1). \tag{1}$$

The destination, in block $i$, first decodes the relay's message $s_i$ which is possible if

$$R_1 < I(X_1; Y). \tag{2}$$

Upon decoding $s_i$, $w_{i-1}$ is restricted to the bin $\mathcal{S}_{s_i}$ which is of the size $2^{n(R-R_1)}$. Since $w_{i-1}$ is encoded by a codebook generated according to $p(x|x_1)$, the destination can successfully decode $w_{i-1}$ in block $i$ if $R$ and $R_1$ satisfy

$$R - R_1 < I(X; Y | X_1). \tag{3}$$

Inequalities (1), (2), and (3) give the decode-and-forward achievable rate for the relay channel:

$$R = \sup_{p(x,x_1)} \min\{I(X, X_1; Y), I(X; Y_1 | X_1)\} \tag{4}$$

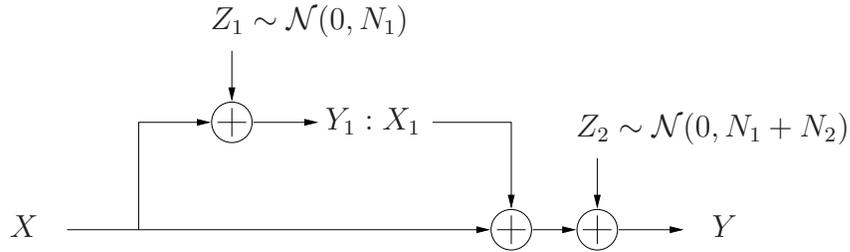

Fig. 3.  The Gaussian relay channel: $X$ is power constrained to $P$; $X_1$ is power constrained to $P_1$.

which is also the capacity if the channel is degraded [1, Theorem 1].

## B. Binning for the Gaussian Relay Channel

This paper focuses on the Gaussian relay channel

$$Y_1 = X + Z_1 \tag{5}$$
$$Y = X + X_1 + Z_2, \tag{6}$$

as shown in Fig. 3, where $Z_1 \sim \mathcal{N}(0, N_1)$ $Z_2 \sim \mathcal{N}(0, N_1 + N_2)$ are independent Gaussian noises. The transmitter has a power constraint $P$. The relay has a power constraint $P_1$. For the decode-and-forward strategy to work, the noise variance at the relay must be smaller than the noise variance at the destination. This channel is not degraded, unless $Z_2$ is a degraded version of $Z_1$ (i.e. $Z_2 = Z_1 + Z'$, where $Z' \sim \mathcal{N}(0, N_2)$ is independent of $Z_1$.)

For the Gaussian relay channel, Cover and El Gamal [1, Section IV] showed that the optimal codebook $\mathcal{X}(\cdot|s_i)$ is additive in the sense that codewords $X(w_i|s_i)$ can be constructed via

$$X(w_i|s_i) = \tilde{X}(w_i) + \sqrt{\frac{(1-\alpha)P}{P_1}} X_1(s_i). \tag{7}$$

where $\tilde{X}(w_i) \sim \mathcal{N}(0, \alpha P)$ and $X_1(s_i) \sim \mathcal{N}(0, P_1)$ are independent codebooks of sizes $2^{nR}$ and $2^{nR_1}$, respectively. This determines an optimal joint distribution $p(x, x_1)$ under the power constraints. The source divides its total power budget $P$ into a fraction $\alpha P$ for transmitting new codeword $w_i$ and a fraction of $(1-\alpha)P$ for cooperatively transmitting the bin index $s_i$ of the previous codeword $w_{i-1}$. The optimal $\alpha$ is determined later.

The additive structure of the $X(w_i|s_i)$ makes practical construction of codes for the Gaussian relay channel feasible. It also makes explicit the fact that $X_1$ and $X$ jointly transmit $s_i$ to the destination at the same time. The decoding process goes as follows. The relay decodes $w_i$ based on $Y_1$.

$$Y_1 = X + Z_1 = \tilde{X}(w_i) + \sqrt{\frac{(1-\alpha)P}{P_1}} X_1(s_i) + Z_1. \tag{8}$$

Since $X_1(s_i)$ is known at the relay, it can be subtracted. Therefore, the successful decoding of $\tilde{X}(w_i)$ is possible if

$$R \leq I(X; Y_1|X_1) = \frac{1}{2}\log\left(1 + \frac{\alpha P}{N_1}\right) \tag{9}$$

The destination observes

$$Y = X + X_1 + Z_2 = \tilde{X}(w_i) + \left(1 + \sqrt{\frac{(1-\alpha)P}{P_1}}\right) X_1(s_i) + Z_2. \tag{10}$$

The decoding of $w_i$ takes place in two stages. First, the decoder decodes $s_i$, the bin index of $w_{i-1}$, while regarding $\tilde{X}(w_i)$ as noise. The decoding is successful if

$$R_1 \leq \frac{1}{2}\log\left(1 + \frac{(\sqrt{P_1} + \sqrt{(1-\alpha)P})^2}{\alpha P + N_1 + N_2}\right). \tag{11}$$

With $s_i$ known, the destination now subtracts $X_1(s_i)$ and proceeds with the decoding of $w_i$ in the second stage. This is done in the next coding block, after the bin index $s_{i+1}$ is decoded. The bin index restricts the candidate codewords into a set of size $2^{n(R-R_1)}$. Thus, decoding is successful whenever

$$R - R_1 \leq I(\tilde{X}; Y|X_1) = \frac{1}{2}\log\left(1 + \frac{\alpha P}{N_1 + N_2}\right). \tag{12}$$

Combining (9) (11) and (12), Cover and El Gamal [1, Theorem 5] derived the following achievable rate for the Gaussian relay channel:

$$R = \max_{\alpha} \min\left\{\frac{1}{2}\log\left(1 + \frac{\alpha P}{N_1}\right), \frac{1}{2}\log\left(1 + \frac{P + P_1 + 2\sqrt{(1-\alpha)PP_1}}{N_1 + N_2}\right)\right\}. \tag{13}$$

The maximizing $\alpha$ in the above expression is $\alpha = 1$ if $P/N_1 \geq P_1/N_2$ which is the case when the optimal strategy is *not* to allocate any portion of the transmitter's power to cooperate with the relay's message. Thus, *no coherent transmission* is needed between the relay and the transmitter. When $P_1/N_2 < P/N_1$, the optimal $\alpha$ is obtained by equating the two rate expressions in (13). This is when coherent transmission of $s_i$ by the source and the relay is beneficial.

The code construction problem to implement the above scheme can be formulated as two subproblems: two codebooks are needed, $\mathcal{X}_1$ of rate $R_1$, and $\tilde{\mathcal{X}}$ of rate $R$. The relay's codebook, $\mathcal{X}_1$, can be constructed as a conventional error-correcting code that guarantees successful decoding at rate $R_1$, at the destination. The source's codebook, $\tilde{\mathcal{X}}$, needs to be constructed so that the relay may decode at

$$\text{SNR}_+ = \frac{\alpha P}{N_1} \quad (14)$$

and the destination may decode under a different SNR

$$\text{SNR}_- = \frac{\alpha P}{N_1 + N_2}, \quad (15)$$

but with the help of extra bin index information at rate $R_1$.

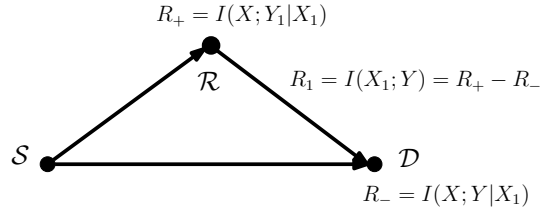

Fig. 4. The code construction problem for decode-and-forward corresponds to two subproblems: constructing a source codebook to simultaneously approach rates $R_+$ and $R_-$, and constructing a conventional relay codebook to approach the rate $R_1 = R_+ - R_-$.

The code construction problem for the Gaussian relay channel is abstracted in a schematic depicted in Fig. 4. The source's codebook should be a capacity-achieving code over an abstracted Gaussian channel, representing the source-relay link, with $\text{SNR} = \text{SNR}_+$ and rate

$$R_+ = \frac{1}{2} \log \left( 1 + \frac{\alpha P}{N_1} \right) \quad (16)$$

which corresponds to (9). The source's codebook, with the help of relay, should also be a capacity-achieving code over an abstracted Gaussian channel, representing the source-destination link, with $\text{SNR} = \text{SNR}_-$ and rate

$$R_- = \frac{1}{2} \log \left( 1 + \frac{\alpha P}{N_1 + N_2} \right). \quad (17)$$

which corresponds to (12).

Note that in the regime where the optimal $\alpha = 1$, the decode-and-forward rate is limited by $R_+$, i.e., extra power $P_1$ at the relay terminal does not improve the overall rate. However, from a code construction point of view, extra power at the relay node simplifies the code construction problem since the relay-destination link can provide more redundancy bits than necessary to the destination[1],[2]. This is the case considered in [16]. In this paper, we focus on the more stringent situation with $R_1 = R_+ - R_-$, which corresponds to the case $P/N_1 \leq P_1/N_2$ with optimum cooperation factor $\alpha < 1$.

Although the coding construction in this section is derived from a full-duplex relay channel, the coding scheme is applicable more generally to *any* relay-destination channel, including a digital relay-destination link and the half duplex relay channel, as long as the relay-destination rate, $R_1$, satisfies $R_1 \geq R_+ - R_-$.

This paper focuses on the practical design of $\tilde{\mathcal{X}}$, while ignoring multiple-access and interference-subtraction issues at the destination. Practical implementation of superposition coding and interference subtraction has been well studied in the literature. The above discussion also ignores error propagation, whereby an incorrect decoding of $X_1(s_{i-1})$ negatively affects the decoding of $\tilde{X}(w_i)$. In practice, the probability of error for the decode-and-forward protocol is bounded by the maximum of failure probabilities of the decoding of the source's message at the relay, the decoding of relay's message at the destination, and the decoding of the source's message at the destination. While error propagation does not impact the design of capacity-achieving codes, it is practically important, especially in term of outage probability in a wireless fading channel.

## C. Bilayer Codes for Parity-Forwarding

A crucial ingredient of the decode-and-forward strategy is binning. How can binning be implemented in practice? If we restrict our attention to Gaussian channels at low SNR (i.e. $R < 1$) for which binary signaling and linear codes are optimum, then binning may be implemented by generating extra parity bits on the codewords of $\tilde{\mathcal{X}}$. The generation of parity bits (or syndromes) is a natural way of partitioning a linear codebook into bins, with codewords in each bin satisfying a particular set of parity equations. The parity bits are exactly the bin indices. The idea of implementing structured binning via syndromes has been used in the past for Slepian-Wolf coding [32] and for channel and source coding with side information [33].

To implement binning and block-Markov coding using this idea, the relay decodes the transmitted codeword $\tilde{X}(w_i)$ in block $i$, and generates extra parity bits for $\tilde{X}(w_i)$, codes them using an independent codebook $\mathcal{X}_1$, and sends the coded bits to the destination in the next block. The destination decodes $\tilde{X}(w_i)$ by utilizing the extra parity bits. Therefore, the decode-and-forward strategy is a *parity-forwarding strategy*.

---

[1] This is shown, for example, in the simulation results of [17], where a smaller gap to Shannon limit is observed when the relay is allowed to transmit for a longer period of time in a half-duplex relay channel.

[2] However, when extra power is available at the relay node, decode-and-forward cannot be the optimal strategy. Higher rate can be obtained by, for example, quantize-and-forward.

We focus on the design of a new LDPC code structure for $\tilde{\mathcal{X}}$ to implement the parity-forwarding strategy. As mentioned earlier, $\mathcal{X}_1$ can be designed as a conventional LDPC code. However, special considerations are needed for the design of $\tilde{\mathcal{X}}$. Let $\tilde{\mathcal{X}}$ be a linear $(n, n-k_1)$ LDPC code with a rate of $(n-k_1)/n$. The codebook $\tilde{\mathcal{X}}$ should be a capacity approaching code for the channel between the source and the relay, i.e., at $\text{SNR}_+$ with a rate approaching (16).

Let $k_2$ be the number of randomly-generated extra parity bits for a source codeword $\tilde{X}(w_i)$ generated by the relay and provided to the destination. Then, $\tilde{X}(w_i)$, should satisfy two sets of parities: $k_1$ zero parities enforced by the source's codebook, and $k_2$ extra presumably nonzero parity bits provided by the relay. Thus, a subcode of $\tilde{\mathcal{X}}$ that satisfies the additional $k_2$ parity checks should form a $(n, n-k_1-k_2)$ capacity-approaching code for decoding at the destination, i.e. at $\text{SNR}_-$ with a rate approaching (17).

The decoding of the subcode of $\tilde{\mathcal{X}}$ with $k_2$ nonzero parity bits can be done in the exact same way as the decoding of a conventional LDPC code. Note that different $k_2$ bin index values correspond to different subcode; they form a coset partition of $\tilde{\mathcal{X}}$ and are related to each other through a coset leader. For linear codes, the subcodes can be generated by the binary addition of the coset leader to the subcode defined by enforcing both $k_1$ and $k_2$ parity bits to zero. Since the subcodes are identical to each other geometrically, we only need to ensure that the subcode represented by the zero-codeword coset leader is well designed. The implementation of the parity-forwarding scheme using LDPC codes is graphically depicted in Fig. 5.

The proposed LDPC code structure is shown in Fig. 6. We call the proposed code structure a *bilayer-expurgated LDPC code*, as the first (lower) layer corresponds to a $(n, n-k_1)$ code (for the source-relay channel) consisting of the $k_1$ zero parity bits and a second (upper) layer consists of the $k_2$ extra parity bits which modifies the first layer in a way that a $(n, n-k_1-k_2)$ subcode represented by the overall graph is suitable for the source-destination channel. The overall graph represents an expurgated subcode of the lower-layer code. Note that the performance of a practical bilayer code is characterized by two gaps to capacity, that at $\text{SNR}_+$ and at $\text{SNR}_-$.

## D. Coding for the Erasure Relay Channel

Before considering the design of bilayer codes for the Gaussian channel, it is useful ask whether such a capacity-achieving bilayer code exists in theory. It is well known that random linear codes are capacity-achieving for the binary symmetric channel (and for the Gaussian channel at a low SNR) under the maximum likelihood decoding. A subcode of a random code is also a random code. Therefore, under the maximum likelihood decoding, a bilayer code can be found to achieve capacities at two different SNRs.

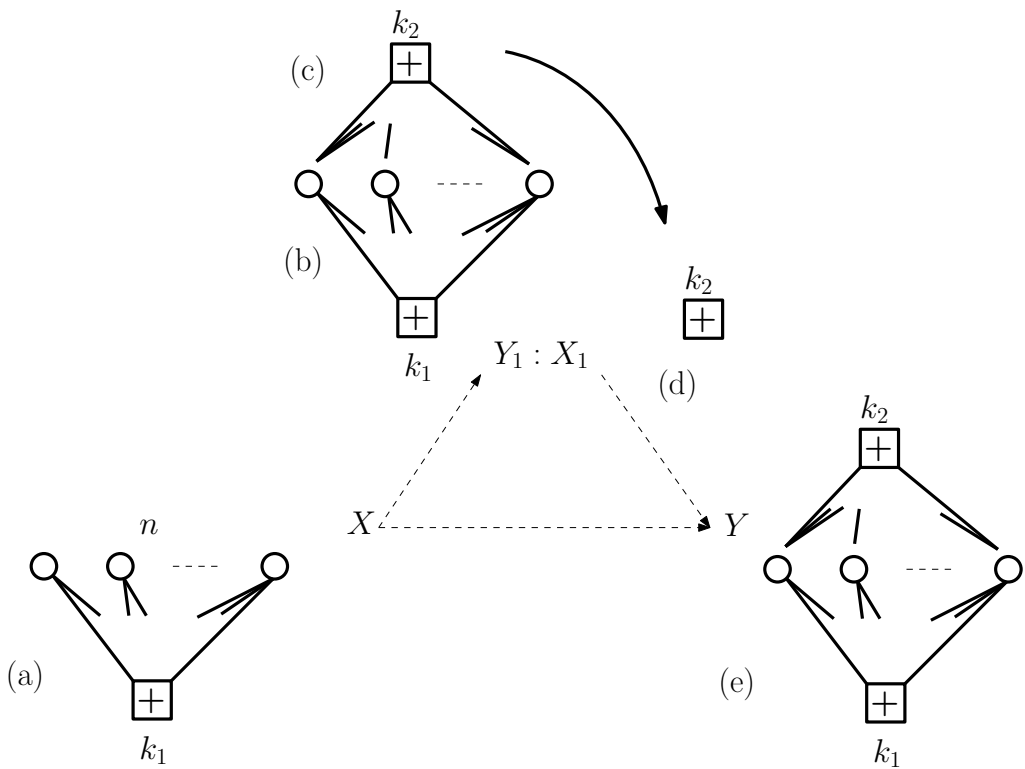

Fig. 5. Parity-forwarding implementation of decode-and-forward using LDPC codes: (a) The source message is encoded using an $(n, n-k_1)$ LDPC code; (b) The relay decodes the source's codeword; (c) The relay then generates $k_2$ extra parity bits; (d) The $k_2$ parity bits are transmitted to the destination using a separate codebook; (e) The destination first decodes the extra $k_2$ parity bits, then decodes the source message over the bilayer code by searching for a codeword that satisfies $k_1$ zero parity bits and $k_2$ nonzero parity bits.

The question becomes more interesting if we consider practical iterative decoding methods. In this realm, theoretical results are available only for the binary erasure channel, for which capacity-achieving degree sequences for low-density parity-check codes under iterative decoding methods have been identified [34] [35] [36].

Consider the binary erasure relay channel as shown in Fig. 7, for which the source-relay channel is a BEC with erasure probability $\epsilon_1$, the source-destination channel is an independent BEC with erasure probability $\epsilon_2 \geq \epsilon_1$, and a relay-destination channel is a digital link with capacity $R_1$. When $R_1 \leq (1-\epsilon_1)-(1-\epsilon_2)$, the capacity of this channel is known to be $C = R_1 + (1-\epsilon_2)$ [37]; decode-and-forward strategy achieves the capacity on this channel.

Can practical codes achieve the capacity for binary erasure relay channel? In a remarkable development motivated by lossy Internet packet transmission applications, Luby [24] showed that instead of using conventional LDPC codes, where codewords satisfying a set of parity constraints are transmitted through the channel, it is possible to devise *universal* low-density generator-matrix (LDGM) codes, termed LT

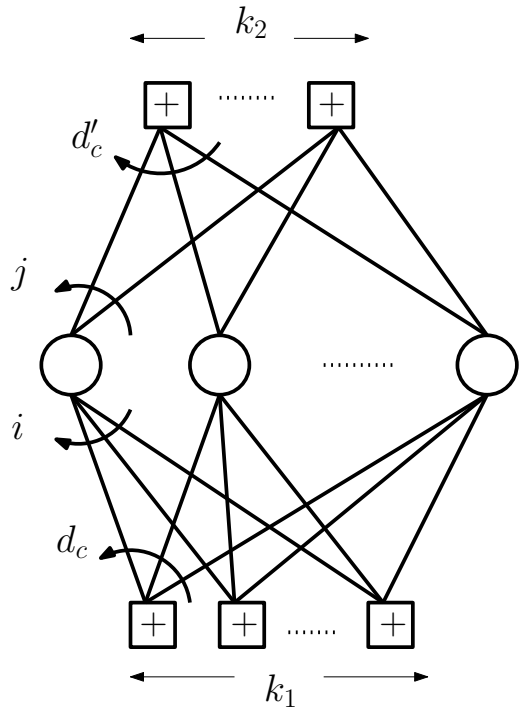

Fig. 6. The bilayer-expurgated code. The lower subgraph represents a LDPC code for source-relay channel. The overall graph represents a LDPC code for the destination.

PSfrag replacements

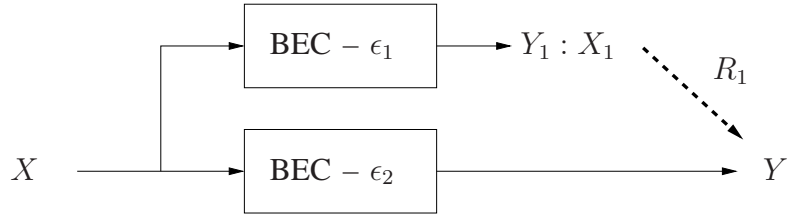

Fig. 7. The binary-erasure relay channel with a digital link from the relay to destination

code, to achieve the BEC capacity for any arbitrary erasure probability. In the LT code construction, random parities generated from a carefully chosen degree distribution are transmitted through the BEC. Luby proved that using a parity generation function of average degree $O(\ln(m))$, one only needs $m + o(m)$ parities to decode the transmitted bit sequence with high probability. Thus, as $m \to \infty$, one can approach the BEC capacity regardless of the erasure probability.

LT codes can be easily adapted to create capacity-achieving codes for the erasure relay channel. Instead of using LT codes as an online code, consider a block code with $m$ source bits and $n + o(n)$ encoded parity bits sent by $X$, with rate $\frac{m}{n} = R_1 + (1 - \epsilon_2)$. Since $R_1 \leq (1 - \epsilon_1) - (1 - \epsilon_2)$, the relay would receive a sufficient number of parities to decode the source bits with a high probability. The relay then independently re-encodes the source bits using the same degree distribution, and sends the additional parities to the destination via the digital link at rate $R_1$. The total number of independent parities at the destination is then $(1 - \epsilon_2)n$ bits from the source plus $R_1 n$ bits from the relay. Thus, decoding would be successful for arbitrary rates below $C = R_1 + (1 - \epsilon_2)$.

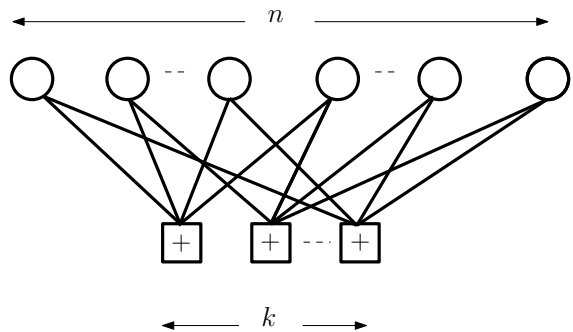

Fig. 8. Schematic diagram of a LDPC code of rate $1 - k/n$ with $n$ variable nodes and $k$ check nodes.

The above argument shows that practical capacity-achieving codes exists for the erasure relay channel. In fact, the above scheme can be further improved in practice by using Raptor codes [25] instead of LT codes to achieve linear-time encoding and decoding performance. However, as mentioned earlier, neither Raptor codes nor LT codes can be used to achieve the Gaussian relay channel capacity, as capacity-achieving rateless codes for general binary symmetric channels have not been found [26]. The rest of the paper focuses on LDPC code design methods for the Gaussian relay channel that come very close to the best achievable decode-and-forward rate.

## III. LDPC CODE PRELIMINARIES

This section reviews the terminologies and design methods for LDPC codes. The design methodology of bilayer LDPC codes for the relay channel is described in the next section and is based on the design procedure described here.

### A. Terminologies

The LDPC code, originally invented by Gallager [22], is a powerful class of linear codes that can approach very close to the Shannon capacity of a Gaussian channel [38]. A LDPC code can be described by a bipartite graph consisting of two sets of nodes: *variable nodes* and *check nodes*. In a binary LDPC code, variable nodes take $\pm 1$ values and a check node requires the module 2 sum of all variable nodes connected to it to be zero. A schematic diagram of a binary LDPC code is shown in Fig. 8, where variable nodes are represented by circles and check nodes are represented by squares.

The *variable degree* of a variable node is the number of edges connected to the node. The *check degree* of a check node is defined similarly. A code is said to have a *regular variable (or check) degree* if the degrees of all variable (or check) nodes are equal. A *regular* LDPC code, originally invented by Gallager, is a LDPC code with both regular variable and regular check degrees. In an *irregular LDPC code* invented in [39], variable nodes or check nodes can have unequal degrees.

An ensemble of irregular LDPC codes is described by two sets of parameters: the *variable degree distribution* $\lambda_i, i \geq 2$, and the *check degree distribution* $\rho_j, j \geq 2$. The variable and check degree distributions define the percentage of *edges* in the graph that are connected to various variable and check degrees. Equivalently, $\lambda_i$ denotes the probability that an edge has a variable degree $i$, $\rho_j$ denotes the probability that an edge has a check degree $j$. Note that $\sum_{i \geq 2} \lambda_i = \sum_{j \geq 2} \rho_j = 1$.

The decoding of a LDPC code is commonly based on a *message passing* algorithm over the edges of the bipartite graph. In a message-passing algorithm, a message corresponds to an edge and represents an estimate for the value of the variable node connected to the edge, possibly along with a soft reliability information for that estimate. Messages are iteratively updated by *variable updates* and *check updates*. Different message-passing algorithms are possible depending on the types of messages and variable and check updates.

In this paper, we focus on the *sum-product* algorithm, also known as the belief propagation algorithm, in which the message corresponding to an edge is the log-likelihood value of the variable node connected to it. (See [2] or [40] for further details.) Associated with such a message is a hard decision. A message is correct/incorrect if its associated hard decision is correct/incorrect. The message-error-probability at a decoding iteration is defined as the percentage of incorrect messages in the graph in that decoding iteration.

The performance of an infinite-length LDPC code can be accurately predicted based on the variable and check degree distributions from which the code ensemble is generated. Density evolution is a tool invented by Gallager [22] and substantially extended by Richardson and Urbanke [2] to analyze the performance of a LDPC code under message-passing decoding. Density evolution tracks the evolution of the message probability density function (PDF) as variable and check updates are performed in successive decoding iterations. Details of density evolution algorithm can be found in [2].

It has been demonstrated that an ensemble of LDPC code exhibits a threshold phenomenon [2]. When decoded over a Gaussian channel with $\text{SNR} > \text{SNR}_{\text{threshold}}$, the decoding error probability arbitrarily approaches zero as the length of the code increases. Conversely, if $\text{SNR} \leq \text{SNR}_{\text{threshold}}$ the decoding error probability is bounded away from zero by a positive constant regardless of the length of the code and number of decoding iterations. The threshold $\text{SNR}_{\text{threshold}}$ is called the *convergence threshold* of the code.

## B. LDPC Code Design

The rate of a LDPC code is determined by the variable and check degree distributions as follows. Let $E$ be the total number of edges in the graph. Then, the total number of variable nodes is given by $E \sum_{i \geq 2} \lambda_i / i$; the total number of check nodes in the graph is given by $E \sum_{j \geq 2} \rho_j / j$. Thus, the rate of the code is given by:

$$R = 1 - \frac{\sum_{j \geq 2} \rho_j / j}{\sum_{i \geq 2} \lambda_i / i}. \tag{18}$$

The LDPC code design problem is to find a pair of variable and check degree distributions that maximize (18) while ensuring successful decoding at a given SNR.

It has been shown that for certain classes of decoding algorithms, the optimum check degree distribution is concentrated around a mean value and a capacity approaching LDPC code has only one or two consecutive check degrees[3] [39, Section 3.3]. As a result, in LDPC design, it is common to fix a regular check degree and optimize the code over the variable degree distribution only. The optimum check degree is often found by trying different values. Empirically, the optimum check degree of the code is found to increase with the maximum allowed variable degree, $\max(d_v)$, and with the rate of the code. Note that the optimal check degree for codes at different rates can be very different. This fact will be important later for the design of bilayer codes, which have to operate at two different rates.

With a fixed check degree, the rate maximization problem (18) is equivalent to the maximization of $\sum_{i \geq 2} \lambda_i / i$. This leads to a linear programming approach to code optimization. Linear programming method for variable-degree optimization first appeared in [39]; the design method was later modified in [38]. The method used in this paper is based on a different approach devised in [18], which is inspired by EXIT-chart based methods. The rest of this section outlines this approach.

Assuming a fixed check-degree distribution, the basic idea is to start with some variable-degree distribution $\lambda_i$, then iteratively improve the overall rate while ensuring convergence by identifying a better $\lambda_i$.

For a fixed $\lambda_i$, the iterative decoding process can be characterized by a set of message PDF at the beginning of each decoding iteration, denoted as $p^l$, $l = 1 \cdots L$, where $L$ is the maximum number of iterations. Here, each decoding iteration is defined to be a check update followed by a variable update for all messages in the graph (i.e. a parallel message-passing schedule is assumed.) The message PDF at beginning the $(l+1)$'th decoding iteration, $p^{l+1}$, can be computed using the density evolution algorithm in [2] or the discretized density evolution scheme of [38]. The message error probability of each iteration

---
[3]See, for example, a database of optimized LDPC codes available at [41].

can be directly calculated from $p^l$.

The main idea is to start with some initial $\lambda_i$ (which determines a set of $p^l$'s), then *assume* that $p^l$'s are fixed and incrementally adjust $\lambda_i$ to maximize the overall rate while ensuring convergence. This incremental adjustment is, of course, not exact, as $p^l$ depends on $\lambda_i$. However, as we shall see, such an assumption yields a low-complexity, yet accurate, code optimization procedure if the incremental adjustment on $\lambda_i$ is sufficiently small, which implies that the change in $p^l$ is also small.

Fixing $p^l$, the incremental adjustment on $\lambda_i$ can be done via linear programming using an approach inspired by EXIT charts [19], [20], [21], [22]. The key ingredient is to define a set of error profile functions an irregular LDPC code, $e_i(p)$, as a function of the input message PDF, for each variable degree $i$ separately, as follows. Consider an auxiliary LDPC code with regular variable degree $i, i \geq 2$, and the same check degree as the irregular code. The degree $i$ error profile, $e_i(p)$, as a function of input message density $p$, is defined as the message error probability after one density-evolution iteration in the auxiliary regular LDPC code, with an initial message PDF $p$. This is closely related to the concept of *elementary EXIT chart* defined in [23]. The difference is that exact density evolution is used; there is no Gaussian approximation of message densities.

Let $e(p^{l+1})$ denote the message error probability corresponding to the message density $p^{l+1}$ in the $(l+1)$'th decoding iteration in the original irregular LDPC code. The degree $i$ error profile, $e_i(\cdot)$, can be used to compute $e(p^{l+1})$ for the irregular code as follows:

$$e(p^{l+1}) = \sum_{i \geq 2} \lambda_i e_i(p^l).$$

The message-passing decoding algorithm converges if the message error probability of the code decreases with each decoding iteration. This can be formulated by a set of convergence inequalities as follows:

$$e(p^{l+1}) = \sum_{i \geq 2} \lambda_i e_i(p^l) < e(p^l), \quad l = 1 \cdots L. \tag{19}$$

The above set of inequalities is linear in $\lambda_i, i \geq 2$, if $p^l$'s are fixed. However, $p^l$'s depend nonlinearly on $\lambda_i$. Nevertheless, (19) can still be used to formulate an approximate linear programming problem to update $\lambda_i$. The idea is to update $\lambda_i$ slowly by enforcing a more stringent convergence condition

$$\sum_{i \geq 2} \lambda_i e_i(p^l) < \mu e(p^l), \quad l = 1 \cdots L \tag{20}$$

where $\mu$ is a *convergence factor* that increases slowly from 0 to 1 in the iterative design process. This works because a small change in $\mu$ corresponds to only a small change in the convergence behavior of the code, and thus a small change in error profile $e_i(p^l)$.

Using (20), an iterative optimization scheme for updating the variable degree distribution can be formulated as follows. A sequence of linear programming problems

$$\max_{\lambda_i, i \geq 2} \sum_{i \geq 2} \lambda_i / i \tag{21a}$$

$$\text{s.t.} \sum_{i \geq 2} \lambda_i e_i(p^l) < \mu^h e(p^l), \ l = 1 \cdots L \tag{21b}$$

$$\sum_{i \geq 2} \lambda_i = 1 \tag{21c}$$

are solved successively, where $h$ denotes the optimization iteration number and $l$ is decoding iteration number. We start with all variable degrees set to $\max(d_v)$, i.e. $\lambda_{\max(d_v)} = 1$, and use this $\lambda_i$ to compute the $e_i(p^l)$ coefficients in (21b). This ensures a small initial $\mu^0$ (as long as an appropriate check degree is selected). We then solve the resulting linear programming program to obtain an updated $\lambda_i$. For this $\lambda_i$, we recompute $e_i(p^l)$, and solve the linear programming problem again with a slightly increased $\mu^h$. The slight increase in $\mu^h$ ensures that the change in $e_i(p^l)$ is small as compared to the previous iteration. The new variable degree distribution, $\lambda_i$, obtained from (21) is then used to update $e_i(p^l)$ coefficients. The optimization is repeated with $\mu^{h+1}$, until $\mu^h$ eventually reaches 1.

This procedure is reminiscent of the EXIT-chart approach, because the value of $\mu$ defines the shape of the convergence behavior. To speed up the $\mu$-update process, we also use a backtracking algorithm: at the end of the $h$'th iteration, a greedy increase in $\mu^h$ is performed. If the resulting degree distribution does not correspond to a converging LDPC code (i.e., (21b) cannot be satisfied with $\mu^{h+1}$), $\mu^{h+1}$ is reduced and the optimization is repeated.

## IV. Design of Bilayer-Expurgated LDPC Codes

We now extend the use of iterative linear programming to design bilayer-expurgated LDPC codes for the relay channel. Toward this end, we first characterize the ensemble of bilayer-expurgated LDPC codes and devise bilayer density evolution as a performance analysis tool appropriate for this new ensemble. Based on bilayer density evolution, the design methodology described in the previous section is then adapted for the optimization of bilayer LDPC codes.

A key simplifying assumption in our linear programming methodology for LDPC code design is that the check degree is concentrated, which is near optimal for the conventional LDPC codes. However, concentrated check degree is difficult to realize for a bilayer-expurgated code, as two sets of parity checks are involved, and the code must work at two different SNRs.

Check-degree optimization is a key aspect of bilayer code design. This paper proposes two bilayer LDPC code designs. The first approach, which is described in this section, assumes two concentrated check degrees at the two sets of parity bits of the bilayer code. The second approach defines a different code ensemble by lengthening a LDPC code and can be thought as a dual of the first approach. The design of bilayer-lengthened codes is described in the next section.

## A. Bilayer-Expurgated LDPC Code Ensemble

The design of bilayer-expurgated LDPC codes is based on a code ensemble defined as follows. The bilayer graph, as shown in Fig. 6, consists of three sets of nodes and two sets of edges. The three sets of nodes correspond to one set of variable nodes, and two sets of check nodes: the *lower check nodes* corresponding to check nodes in the lower subgraph of Fig. 6, and the *upper check nodes* corresponding to check nodes in the upper subgraph in Fig. 6. Edges are grouped in two sets: those connecting the variable nodes to the lower check nodes, and those connecting variable nodes to the upper check nodes. We call an edge a *lower edge*, if it connects a variable node to a lower check node. Similarly, an *upper edge* denotes an edge belonging to the upper subgraph in Fig. 6.

The lower degree of a variable node is defined as the number of lower edges connected to it. Likewise, the upper degree of a variable node is the number of upper edges connected to it. The lower degree of an *edge* is defined as the lower degree of the variable node it is connected to, and similarly the upper degree of an edge is the upper degree of the variable node connected to that edge. The minimum lower variable degree is 2 as the lower subgraph should be a valid LDPC code for the source-relay channel. The minimum upper variable degree is 0, since some variable nodes may not participate in any of the $k_2$ extra parity checks generated by the relay. A variable node is said to have degree $(i,j)$ if it has a lower degree $i$ and an upper degree $j$. Similarly, an edge is of degree $(i,j)$ if it is connected to a degree $(i,j)$ variable node.

We assume regular check degrees for check nodes in the lower and upper graphs. The lower check degree of a bilayer graph, $d_c$, denotes the number of edges connected to check nodes in the lower subgraph. Likewise, the upper check degree, $d'_c$, equals to the number of edges connected to an upper check node.

The ensemble of bilayer LDPC codes can be characterized by a variable degree distribution, $\lambda_{i,j}, i \geq 2, j \geq 0$, which defines the percentage of edges with lower degree $i$ and upper degree $j$ and a parameter $\eta$ which defines the percentage of lower edges in the bilayer graph. In other words, the probability that an edge is connected to a variable with lower degree $i$ and upper degree $j$ is given by $\lambda_{i,j}$, and the probability that an edge is a lower edge is given by $\eta$. Note that $\sum_{i\geq 2, j\geq 0} \lambda_{i,j} = 1$, and $0 < \eta < 1$.

Note also that a bilayer LDPC code reduces to a conventional LDPC code if $\lambda_{i,j} = \lambda_i \lambda_j, i \geq 2, j \geq 0$ for some set of parameters $\lambda_i$ with $\lambda_0 = \lambda_1 = 0$.

## B. Bilayer Density Evolution

Because the ensemble of bilayer-expurgated LDPC codes is statistically different from a conventional LDPC code ensemble, conventional density evolution algorithm must be modified in order to accurately predict the performance of the bilayer code. In the conventional density evolution analysis, the input message densities to all check nodes at each density evolution iteration are the same, since the probability that an edge emanating from a check node is connected to a degree $i$ variable node is *equal* to $\lambda_i$ for *all* check nodes. However, in a bilayer-expurgated code there is a distinction between lower edges and upper edges. Therefore, in order to predict the performance of a bilayer code, evolution of two densities should be tracked: the lower density corresponding to the density of messages in the lower subgraph, and the upper density corresponding to the density of messages in the upper subgraph.

Let $p^l$ and $q^l$ denote the message densities at the input of lower and upper check nodes in the lower and upper subgraphs, respectively, at the beginning of the $l$'th decoding iteration. The message densities after a check update can be computed for $p^l$ and $q^l$ using the conventional density evolution check update as described in [2]. Let $p'^l$ and $q'^l$ denote the evolved versions of $p^l$ and $q^l$ after the check updates. For log-likelihood message-passing decoding, the density-evolution update at a degree $(i,j)$ variable node can be computed from $p'^l$ and $p'^l$ to obtain the message densities, $p^{l+1}_{i,j}$ and $q^{l+1}_{i,j}$, as follows:

$$p^{l+1}_{i,j} = \left(\otimes^{i-1} p'^l\right) \otimes \left(\otimes^j q'^l\right) \otimes p_c, \quad i \geq 2, j \geq 0 \qquad (22)$$

$$q^{l+1}_{i,j} = \left(\otimes^i p'^l\right) \otimes \left(\otimes^{j-1} q'^l\right) \otimes p_c, \quad i \geq 2, j \geq 1 \qquad (23)$$

where $p_c$ denotes the density of the log-likelihood ratio received over the channel, and $\otimes^i$ denotes convolution of order $i$. (By convention, for any density $f$, $\otimes^1 f = f$ and $\otimes^0 f = \delta$, where $\delta$ denotes the Dirac delta function.) The input message densities to lower and upper check nodes, at the beginning of the $(l+1)$'th iteration can be computed as follows:

$$p^{l+1} = \sum_{i\geq 2, j \geq 0} \frac{i}{i+j} \lambda_{i,j} p^{l+1}_{i,j} \qquad (24)$$

$$q^{l+1} = \sum_{i \geq 2, j \geq 0} \frac{j}{i+j} \lambda_{i,j} q_{i,j}^{l+1} \tag{25}$$

Note that the probability that a degree $(i,j)$ edge is a lower edge is given by $i/(i+j)$.

Similar to the error profile function for a conventional LDPC code, the lower-graph degree $(i,j)$ error profile function, $e_{i,j}^1(p^l, q^l)$, is defined for a bilayer-expurgated LDPC code as the message error probability corresponding to the density $p_{i,j}^{l+1}$, after one density evolution iteration with input message densities $p^l$ and $q^l$. Similarly, $e_{i,j}^2(p^l, q^l)$ is defined as the message error probability corresponding to $q_{i,j}^{l+1}$ after one density evolution iteration for input message densities $p^l$ and $q^l$. Let $e(p^{l+1}, q^{l+1})$ denote the overall message error probability in the bilayer graph corresponding to the message densities $p^{l+1}$ and $q^{l+1}$. The overall message error probability at the beginning of the $(l+1)$'th decoding iteration, $e(p^{l+1}, q^{l+1})$, can be computed as a linear combination of $e_{i,j}^1(p^l, q^l)$ and $e_{i,j}^2(p^l, q^l)$ as follows:

$$e(p^{l+1}, q^{l+1}) = \sum_{i \geq 2, j \geq 0} \lambda_{i,j} \left( \frac{i}{i+j} e_{i,j}^1(p^l, q^l) + \frac{j}{i+j} e_{i,j}^2(p^l, q^l) \right). \tag{26}$$

The above formulation allows an approximate linear programming optimization of $\lambda_{i,j}$.

## C. Bilayer-Expurgated LDPC Code Optimization

The design of a bilayer-expurgated LDPC code involves finding a variable degree distribution $\lambda_{i,j}, i \geq 2, j \geq 0$, a parameter $\eta$, and a pair of check degrees, $d_c$ and $d_c'$, such that the lower subgraph represents a capacity-approaching LDPC code over the source-relay channel at $\text{SNR}_+$, and the overall bilayer code is capacity approaching at $\text{SNR}_- < \text{SNR}_+$.

One way to formulate the design problem is to fix $d_c$, $d_c'$, and jointly optimize $\lambda_{i,j}$ and $\eta$. This approach is taken in our previous work [42]; it is equivalent to a joint optimization of both the lower subgraph and the overall graph to achieve the highest overall rate. In this paper, we utilize a different approach by *fixing* the lower-graph code to be a capacity-approaching LDPC code at $\text{SNR}_+$ and searching for a variable degree distribution, $\lambda_{i,j}$, that is consistent with the lower-graph code and is capacity approaching at $\text{SNR}_-$.

We formulate the rate maximization problem for the overall code as follows. Fixing the check degrees $d_c$, $d_c'$, the rate of the bilayer graph is related to the parameter $\eta$, since $\eta$ depends on the number of check nodes in the graph

$$\eta = \frac{d_c k_1}{d_c k_1 + d_c' k_2}. \tag{27}$$

By fixing the lower graph, i.e., fixing $n$, $k_1$ and the lower variable degree distribution $\lambda_i$, the rate of the bilayer code, defined by $1 - (k_1 + k_2)/n$, can be maximized by minimizing $k_2$ or equivalently maximizing $\eta$. The distribution $\lambda_i$ is related to $\lambda_{i,j}$ as follows:

$$\lambda_i = \frac{1}{\eta} \sum_{j \geq 0} \frac{i}{i+j} \lambda_{i,j}. \tag{28}$$

For a fixed $\lambda_i$, (28) can be rewritten in a linear format in terms of $\lambda_{i,j}$ and $\eta$:

$$\sum_{j \geq 0} \frac{i}{i+j} \lambda_{i,j} - \eta \lambda_i = 0. \tag{29}$$

Fixing $d_c$, $d_c'$, and $\lambda_i$, an approximate linear programming update for $\lambda_{i,j}$ and $\eta$ can be formulated using (26) to iteratively maximize $\eta$ as follows:

$$\max_{\lambda_{i,j}, \eta} \eta \tag{30a}$$

$$\text{s.t.} \quad \sum_{j \geq 0} \frac{i}{i+j} \lambda_{i,j} - \eta \lambda_i = 0 \qquad i \geq 2 \tag{30b}$$

$$\sum_{i \geq 2, j \geq 0} \lambda_{i,j} \left( \frac{i}{i+j} e^1_{i,j}(p^l, q^l) + \frac{j}{i+j} e^2_{i,j}(p^l, q^l) \right) < \mu^h e(p^l, q^l), \qquad l = 1 \cdots L \tag{30c}$$

$$\sum_{i \geq 2, j \geq 0} \lambda_{i,j} = 1 \tag{30d}$$

where $h$ is the optimization iteration number, and $l$ is the decoding iteration number. The coefficient $0 < \mu^h < 1$ plays the same role as the $\mu^h$ in (21) and is slightly increased at each optimization iteration toward 1. The error profiles $e^1_{i,j}(p^l, q^l)$, $e^2_{i,j}(p^l, q^l)$ and $e(p^l, q^l)$ are recomputed at the end of each optimization iteration using bilayer density evolution, given the new $\lambda_{i,j}$ and $\eta$.

To initialize the above iterative optimization, an initial degree distribution, $\lambda_{i,j}$, that is consistent with the lower-graph degree distribution in terms of (29) and guarantees a fast decoding convergence with a small $\mu^0 > 0$ should be found. Such a degree distribution can be found using a linear programming optimization that *minimizes* $\eta$, since minimizing $\eta$ or equivalently maximizing $k_2$ corresponds to adding as many extra parity bits as possible which ensures a fast decoding convergence. The initializing linear-programming optimization can be cast as follows:

$$\min_{\lambda_{i,j},\eta} \quad \eta \tag{31a}$$

$$\text{s.t.} \quad \sum_{i \geq 2, j \geq 0} \lambda_{i,j} = 1 \tag{31b}$$

$$\sum_{j \geq 0} \frac{i}{i+j} \lambda_{i,j} - \eta \lambda_i = 0 \qquad i \geq 2 \tag{31c}$$

To complete the design methodology of bilayer LDPC codes, we need to pick appropriate check degrees $d_c$ and $d'_c$. Unfortunately, the overall bilayer code cannot always have a concentrated check degree (i.e. $d_c = d'_c$), as one would like. The reason is that the optimum average check degrees at $\text{SNR}_+$ and $\text{SNR}_-$ can be far apart if the gap between $\text{SNR}_+$ and $\text{SNR}_-$ is large. In this case, the upper check degree, $d'_c$, should be small enough to compensate the effect of a large lower check degree, $d_c$, in order to lower the average check degree of the bilayer code.

An appropriate check-degree pair, $d_c$ and $d'_c$, can be found for a bilayer code by an exhaustive search over a reasonable range of values for $d_c$ and $d'_c$. In our scheme, we pick a check degree $d_c$ and find an optimized conventional LDPC code corresponding to the lower-graph code with regular check degree $d_c$ at $\text{SNR}_+$, using the design scheme described in Section III-B. Then, we try the optimization procedure of (30) for various values of $d'_c$ to find a suitable $d'_c$. In some cases, this procedure needs to be repeated several times to find a satisfactory $d_c$ and $d'_c$ pair.

As mentioned earlier, the optimum check degree for a conventional LDPC code is often concentrated around a fixed value. Thus, when the gap between $\text{SNR}_+$ and $\text{SNR}_-$ is small, the difference between the optimal $d_c$ and $d'_c$ is likely to be small, and this scheme works well. However, if the gap between $\text{SNR}_+$ and $\text{SNR}_-$ is large, the optimal check degree $d'_c$ is often much smaller than $d_c$, resulting a larger gap to capacity. However, in the extreme case, the optimal $d'_c$ may become 1, in which case a new code structure emerges, and a new code design methodology is called for.

Fig. 9 shows the effect of $d'_c = 1$ on the structure of the bilayer-expurgated graph. The lower graph in this case is split into two parts: variable nodes connected to the upper checks on the left, and all other variable nodes on the right. A regular check degree $d'_c = 1$, in effect, removes the variable nodes in the upper part of the graph, by completely determining their values. This completely changes the structure of the bilayer graph. In the next section, we consider the code ensemble corresponding to this new graph, and develop new design methodology and the analysis tools for it.

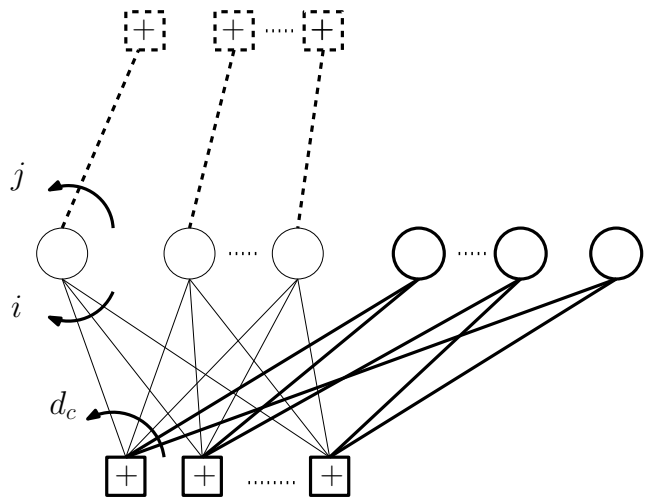

Fig. 9. The effect of $d'_c = 1$ on the bilayer-expurgated graph of Fig. 6. The lower graph is *split* into two sections.

## V. BILAYER-LENGTHENED LDPC CODE

By removing the upper edges in Fig. 9, the variable nodes of a bilayer-expurgated LDPC code are split into two parts. In this case, it is natural to consider a new bilayer LDPC code ensemble, in which different variable-degree distributions are assigned to the two groups of variable nodes. This corresponds to a code structure for which the overall lower graph of Fig. 9 must be capacity-achieving at $SNR_-$, while the right part of the lower graph must be capacity-achieving at $SNR_+$. This new code is schematically shown in Fig. 10. The overall lower code can be thought of as a *lengthened* version of the right part of the lower code. This new ensemble is named bilayer-lengthened LDPC code in this paper.

The parity-forwarding scheme with this new bilayer codes can be described as follows. The source encodes its data using the bilayer code corresponding to the overall graph shown in Fig. 10. Thus, each codeword satisfies all parity-check nodes present in the bilayer graph (in contrast to the earlier bilayer code in which the source encodes its data over the lower subgraph). The relay decodes the source's data over the bilayer graph and forwards the values of the upper variable nodes to the destination, using a separate codebook. (Note that the relay re-encodes variable bits, whereas, in the previous structure the relay re-encodes parity bits.) The destination first *removes* the upper part of the graph, then updates the value of parity-check nodes in the graph. For example, the new value of a parity-check node corresponding to the constraint $v_1 + v_2 + v_3 + v_4 = 0$ after removing $v_1$ and $v_2$ would be $v_1 + v_2$ corresponding to the constraint $v_3 + v_4 = v_1 + v_2$. (The new values of check nodes play the role of a bin index for the received codeword at the destination.) Finally, the destination decodes the remainder of the codeword over the *lower subgraph* (in contrast to the earlier bilayer code in which the destination decodes the source's codeword over the overall graph).

The advantage of this scheme is that the check degrees are reduced after the removal of the upper graph. Therefore, this code is suitable for a relay channel with a large gap between the $\text{SNR}_+$ and $\text{SNR}_-$.

Many of features of the bilayer-lengthened LDPC code are the dual of the bilayer-expurgated code: the roles of variable nodes and check nodes are interchanged in the parity-forwarding scheme; the source encodes its data over the lower graph in one, and over the overall graph in the other. The bilayer-expurgated code performs well for small gap of $\text{SNR}_+$ and $\text{SNR}_-$; the bilayer-lengthened code works well for larger gaps.

The bilayer-expurgated code is closely related to rate-compatible LDPC codes for HARQ, which are often devised by randomly puncturing a high-rate code to produce low-rate codes (e.g. [29], [27], [31]). The design methodology proposed in this paper differs from random puncturing, as the degree distributions of the punctured bits and the remaining bits are explicitly designed, as shown in the next subsection. The bilayer-lengthened code structure considered in this paper is inspired by a code construction, called Matrioshka codes, introduced in [43] for the universal Slepian-Wolf source coding problem.

## A. Bilayer-Lengthened LDPC Code Ensemble

Similar to the bilayer-expurgated code, the bilayer-lengthened graph consists of three sets of nodes and two sets of edges. The nodes are grouped into one set of check nodes (in contrast to the earlier bilayer graph in which there is one set of variable nodes), and two sets of variable nodes (in contrast to the earlier bilayer graph in which there are two sets of check nodes): the *lower variable nodes* corresponding to the variable nodes in the lower subgraph of Fig. 10, and the *upper variable nodes* corresponding to the variable nodes in the upper subgraph in Fig. 10. The edges are grouped in two sets: those connecting the check nodes to the lower variable nodes, and those connecting check nodes to the upper variable nodes. We call an edge a *lower edge*, if it connects a check node to a lower variable node. Similarly, an *upper edge* denotes an edge belonging to the upper subgraph in Fig. 10.

Each check node in the bilayer-lengthened graph has $d_c$ edges in the lower subgraph and $d'_c$ edges in the upper subgraph. Similar to a conventional LDPC code, the degree of a variable node is defined as the number of edges connected to it. An edge is said to have a variable degree $i$ if it is connected to a variable node of degree $i$.

The ensemble of bilayer-lengthened LDPC codes is defined by the *lower variable degree distribution*, the *upper variable degree distribution*, and two regular check degrees $d_c$ and $d'_c$. The lower variable degree distribution, $\lambda^1_i, i \geq 2$, defines the percentage of lower edges of various degrees in the lower subgraph, i.e., the probability that a lower edge is connected to a degree $i$ variable node is given by $\lambda_i^1$. Similarly, the upper variable-degree distribution, $\lambda_i^2, i \geq 2$, describes the probability that an upper edge is of degree $i$. The lower and upper distributions, $\lambda_i^1$ and $\lambda_i^2$, satisfy $\sum_{i \geq 2} \lambda_i^1 = 1$, and $\sum_{i \geq 2} \lambda_i^2 = 1$.

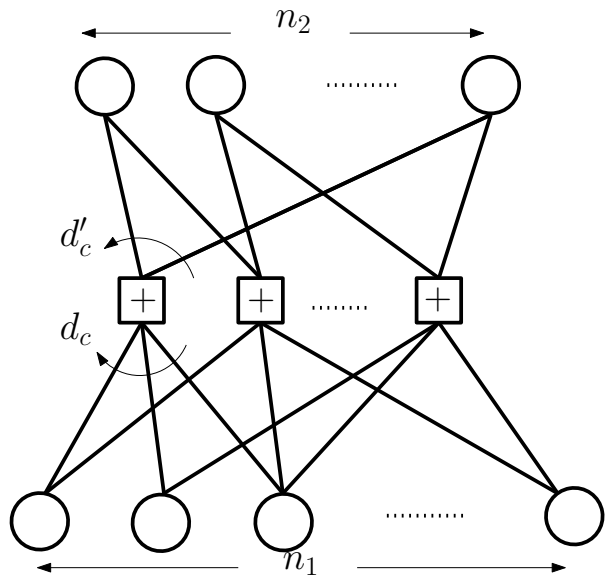

Fig. 10. The bilayer-lengthened LDPC code. The relay decodes the overall code and provides the value of upper variable nodes to the destination, using a separate codebook. The destination decodes the lower subgraph.

Note that the ensemble of bilayer-lengthened LDPC codes is not equivalent to either the conventional LDPC codes or the bilayer-expurgated LDPC codes discussed earlier, because in both of these earlier code ensembles, the variable degree distributions for all variable nodes are the same. Hence, density evolution tools for conventional LDPC codes and for bilayer-expurgated codes are not valid for the bilayer-lengthened LDPC code and should be modified.

## B. Bilayer Density Evolution

The densities of messages over lower and upper edges are in general not equal at each decoding iteration in a bilayer-lengthened LDPC code. This is because the lower and upper edges have different variable degree distributions. Thus, similar to the case of the bilayer-expurgated LDPC codes, to predict the performance of an infinite-length bilayer-lengthened LDPC code, we need to track the evolutions of *two* densities in the upper and lower subgraphs of the lengthened graph.

Let $p^l$ and $q^l$ denote the message densities in the lower and upper parts of the graph at the beginning of the $l$'th decoding iteration. Let $p'^l$ and $q'^l$ denote the evolved version of $p^l$ and $q^l$ after check updates. Let $\oplus$ denote the check density-update operation as described in [2], e.g., $f = f_1 \oplus f_2$ is the output message density after an update at a check node of degree 3. Then, the output message density at a check node of degree $d$ with input message density $f$ can be computed as $\oplus^{d-1} f = f \oplus f \oplus \cdots \oplus f$. Hence, $p^{\prime l}$ and $q^{\prime l}$ can be computed using the check density-update operation as follows:

$$p^{\prime l} = (\oplus^{d_c-1} p^l) \oplus (\oplus^{d'_c} q^l), \qquad d_c > 1 \tag{32a}$$

$$q^{\prime l} = (\oplus^{d_c} p^l) \oplus (\oplus^{d'_c-1} q^l), \qquad d'_c > 1 \tag{32b}$$

and for $d'_c = 1$:

$$p^{\prime l} = (\oplus^{d_c-1} p^l) \oplus (q^l), \qquad d_c > 1 \tag{32c}$$

$$q^{\prime l} = (\oplus^{d_c} p^l) \tag{32d}$$

where $\oplus^1 f \triangleq f$ for any density $f$.

The computation of variable density updates is straightforward using the convolution operation. Let $p_i^{l+1}$ denote the output message density after a variable update at a variable node of degree $i$ in the lower subgraph, with an input message density $p^{\prime l}$. Let $q_i^{l+1}$ denote the output message density after a variable update at a variable node of degree $i$ in the upper subgraph, with an input message density $q^{\prime l}$. Using the convolution operation $\otimes$, we have

$$p_i^{l+1} = \otimes^{i-1} p^{\prime l} \otimes p_c, \qquad i \geq 2 \tag{33}$$

$$q_i^{l+1} = \otimes^{i-1} q^{\prime l} \otimes p_c, \qquad i \geq 2 \tag{34}$$

where $p_c$ is the channel message density.

The message densities in the lower and upper subgraphs after the variable update (i.e. at the beginning of $(l+1)$'th decoding iteration), $p^{l+1}$ and $q^{l+1}$, can be computed from $p_i^{l+1}$ and $q_i^{l+1}$ as follows:

$$p^{l+1} = \sum_{i \geq 2} \lambda_i^1 p_i^{l+1} \tag{35}$$

$$q^{l+1} = \sum_{i \geq 2} \lambda_i^2 q_i^{l+1}. \tag{36}$$

Let $e(p^{l+1}, q^{l+1})$ denote the message error probability of the message densities $p^{l+1}$ and $q^{l+1}$ at the beginning of the $(l+1)$'th decoding iteration. Let $e_i^1(p^l, q^l)$ denote the message error probability corresponding to $p_i^{l+1}$, which is the message density of degree-$i$ lower nodes after one density evolution iteration with input message densities $p^l$ and $q^l$. Similarly, let $e_i^2(p^l, q^l)$ denote the message error probability corresponding to $q_i^{l+1}$, which is the message density of degree-$i$ upper nodes after one density evolution iteration with input message densities $p^l$ and $q^l$. The overall message error probability at the beginning of the $(l+1)$'th iteration, $e(p^{l+1}, q^{l+1})$, can be found as a linear combination of $e_i^1(p^l, q^l)$ and $e_i^2(p^l, q^l)$ functions as follows:

$$e(p^{l+1}, q^{l+1}) = \sum_{i \geq 2} \eta \lambda_i^1 e_i^1(p^l, q^l) + (1-\eta) \lambda_i^2 e_i^2(p^l, q^l). \tag{37}$$

where $\eta = d_c/(d_c + d_c')$ denotes the percentage of lower edges in the bilayer-lengthened graph. The approximate linear structure of (37) is used to form an iterative linear programming procedure to update the variable-degree distributions $\lambda_i^1$ and $\lambda_i^2$ as discussed in the next subsection.

## C. Bilayer-Lengthened LDPC Code Optimization

The design of a bilayer-lengthened LDPC code involves finding a pair of variable degree distributions, $\lambda_i^1$, $\lambda_i^2$ ($i \geq 2$), and a pair of check degrees, $d_c$ and $d_c'$, for the lower and upper subgraphs in the bilayer structure of Fig. 10, such that the overall graph is a capacity-approaching LDPC code for a Gaussian channel at $\text{SNR}_+$, while the lower graph is a capacity-approaching LDPC code at $\text{SNR}_-$.

Similar to the previous design, we fix the check degrees $d_c$ and $d_c'$. (Appropriate check degrees, $d_c$ and $d_c'$, can be found by an exhaustive search over a reasonable range of values for $d_c$ and $d_c'$.) We also fix the lower variable-degree distribution $\lambda_i^1$ to be a capacity-approaching distribution for a conventional LDPC code optimized at $\text{SNR}_-$, (which is found independently.) The design problem is now reduced to finding an upper variable-degree distribution $\lambda_i^2$, such that the overall lengthened graph represents a capacity-approaching code at $\text{SNR}_+$. (Note that in contrast to the design problem of a bilayer-expurgated code, the lower-rate code is fixed here, and the higher-rate code is optimized.)

The rate of the overall bilayer-lengthened code is $1 - k/(n_1 + n_2)$, where $k$ denotes the number of check nodes, $n_1$ is the number of lower variable nodes, and $n_2$ is the number of upper variable nodes. The number of upper variable nodes, $n_2$, is given by $d_c' k \sum_{i \geq 2} \lambda_i^2/i$. Thus, fixing the lower-graph code and $d_c'$, the rate of the overall graph can be maximized by maximizing $\sum_{i \geq 2} \lambda_i^2/i$. To ensure convergence of the overall code, we make use of the error profile function (37). More specifically, fixing $\eta$, $d_c$, and $d_c'$, the linear programming update for $\lambda_i^2$ can be formulated as follows:

$$\max_{\lambda_i^2} \sum_{i \geq 2} \lambda_i^2/i \tag{38a}$$

$$\text{s.t.} \sum_{i \geq 2} \eta \lambda_i^1 e_i^1(p^l, q^l) + (1-\eta) \lambda_i^2 e_i^2(p^l, q^l) < \mu^h e(p^l, q^l), \quad l = 1 \cdots L \tag{38b}$$

$$\sum_{i \geq 2} \lambda_i^2 = 1 \tag{38c}$$

where $h$ denotes the optimization iteration round, and $l$ is the decoding iteration number. The new upper variable-degree distribution is used to update the coefficients $e_i^1(p^l, q^l)$, $e_i^2(p^l, q^l)$, and $e(p^l, q^l)$ for the next optimization round through bilayer density evolution. The coefficient $\mu^h$ is slowly increased toward 1. This enforces an approximate local linearity condition with respect to $\lambda_i^2$, in the same way as in (20). As an initialization value for $\lambda^2$, we set $\lambda^2_{\max(d_v)} = 1$.

The bilayer-lengthened LDPC code is a suitable code structure, if the gap between $\text{SNR}_+$ and $\text{SNR}_-$ is large. However, it has larger gaps to the capacity for smaller values of $\text{SNR}_-$ and smaller differences between $\text{SNR}_+$ and $\text{SNR}_-$. This is because, for a fixed subgraph the minimum number of upper variable nodes is given by $k/\max(d_v)$, which corresponds to an upper subgraph with regular variable degree $\max(d_v)$ and with $k$ check nodes each with degree $d_c' = 1$. Thus, a small $\text{SNR}_-$ (which implies a large $k$), leads to a large minimum additional variable nodes needed in the lengthening process, and consequently a larger minimum $\text{SNR}_+ - \text{SNR}_-$. However, by increasing $\max(d_v)$, the minimum $\text{SNR}_+ - \text{SNR}_-$ can be reduced.

When the gap between $\text{SNR}_+$ and $\text{SNR}_-$ is small, the bilayer-expurgated LDPC code design of Section IV has good performance. In fact, the rate difference can be arbitrarily small for the bilayer-expurgated code. Thus, the bilayer-expurgated LDPC code and the bilayer-lengthened LDPC code are complementary structures that cover the entire range of rates/SNRs.

## VI. Code Construction and Numerical Results

Using the described schemes, six codes, listed in Tables I-VI, are designed for binary-input Gaussian channels with various noise parameters. The maximum variable degree, $\max(d_v)$, for all cases, is chosen to be 20. To speed up the design procedure, for both bilayer-expurgated and bilayer-lengthened LDPC codes, discretized density evolution approach of [38] is utilized with 13-bit quantization and a maximum log-likelihood value 25. To verify the asymptotic infinite-length threshold, the empirical bit-error-rate (BER) performance curves for practically constructed codes are shown in Fig. 11. The block lengths for bilayer-expurgated codes are in the order of 100,000. The block lengths for bilayer-lengthened codes are 70,000.

Code A is constructed for a small rate difference $R_+ - R_-$, using the bilayer-expurgated structure. Code B is constructed using the bilayer-lengthened structure, which is more suitable for a large rate difference. Codes C and D are designed to compare the performance of expurgated and lengthened structures at low SNRs for target rates $0.3$ and $0.4$. For medium SNRs, Codes E and F are designed using the expurgating and lengthening structures for target rates $0.5$ and $0.7$. We observe that for most rate pairs the lengthened structure outperforms the expurgated structure slightly. The bilayer-expurgated structure has a better performance for target rates that are very close to each other and at low SNRs. Over a wide range of SNRs, the asymptotic infinite-length threshold obtained is at most 0.24dB away from the theoretical limit, while finite-length BER results are within at most 0.6dB of the capacity.

More specifically, Code A (Table I) is designed for target rates $0.65$ and $0.7$. Using the bilayer-expurgated structure, this code achieves the rate pair $R_- = 0.6363$ and $R_+ = 0.7000$, i.e., the achieved rate is less than 2.1% smaller than the target $R_-$ rate. (The lower graph, corresponding to the higher-rate code, is fixed to be a conventional LDPC code with rate 0.7 from [41].) The best check-degree pair for this code is found by exhaustive search to be $d_c = 15$ and $d'_c = 8$. The convergence threshold of the overall bilayer code, as predicted by bilayer density evolution, is within a 0.1727 dB gap to the theoretical limit. At BER $= 10^{-4}$, the SNR gap to the Shannon limit of the lower rate channel is about $0.33$ dB at a block length of 100,000.

At a large SNR differences, the bilayer-expurgating LDPC code does not show a good performance. Code B (Table II) is designed for a large SNR difference of about 9 dB, using the bilayer-lengthened structure, for target rates $0.3$ and $0.9$. The achieved rate pair using the bilayer-lengthened structure is $R_- = 0.2871$ and $R_+ = 0.8932$, which are less than 4.2% and 0.75% below the target rates. The convergence thresholds, as predicted by the density evolution, are within a 0.2369 dB gap to the lower-rate channel capacity and within a 0.1357 dB gap to the higher-rate channel capacity. At a BER of $10^{-4}$, the corresponding SNR is within less than 0.25 dB gap to the Shannon limit for a block length of 70,000. (The lower rate component is designed using the scheme described in Section III-B.)

To compare the performance of the bilayer-expurgated structure and the bilayer-lengthened structure, Codes C and Code D (Tables III and IV) are constructed using the two structures for target rates $0.3$ and $0.4$. The achieved rates using the expurgating structure are 8.2% and 3.6% below the target rates $R_+$ and $R_-$, respectively. The achieved rates using the lengthening structure outperforms those of the expurgating structure and are 3.45% and 3.9% below the target rates $R_+$ and $R_-$, respectively. At BER $= 10^{-4}$, the SNR gap to the capacity of the lower rate channel for Code C is close to 0.9 dB for a block length of 100,000, while Code D has a less than 0.6 dB gap to the Shannon limit for a block size of 70,000. (The higher-rate component of the expurgating structure and the lower-rate component of the lengthening structure are fixed with conventional LDPC codes designed using the scheme described in Section III-B.)

Finally, Codes E and F (Tables V and VI) are designed to compare the performance of the proposed expurgating and lengthening schemes at target rates $0.5$ and $0.7$. The achievable rates of the bilayer-

32TABLE I

BILAYER-EXPURGATED LDPC CODE FOR SNR$_+$=2.7330 DB SNR$_-$=1.9262 DB

| Bilayer-Expurgated LDPC code (A) | | | | |
|---|---|---|---|---|
| | $\lambda_{i,0}$ | $\lambda_{i,1}$ | $\lambda_{i,3}$ | $\lambda_{i,4}$ |
| $i=2$ | 0.1398 | 0.0408 | 0 | 0 |
| $i=3$ | 0.1323 | 0.0885 | 0 | 0 |
| $i=6$ | 0.0831 | 0 | 0 | 0 |
| $i=7$ | 0.0295 | 0.1332 | 0 | 0 |
| $i=20$ | 0 | 0 | 0.2600 | 0.0928 |
| $\eta = 0.8982$, $d_c = 15$, $d'_c = 8$ | | | | |
| $R_- = 0.6363$, Threshold gap=0.1727 dB | | | | |
| $R_+ = 0.7000$, Threshold gap=0.08474 dB | | | | |

expurgated Code E are 7.62% and 0.0% below the target rates $R_+$ and $R_-$, respectively. As a comparison, the bilayer-lengthened Code F achieves a rate pair within 2.46% and 1.3% below the target rates $R_+$ and $R_-$. Asymptotically, the convergence threshold of the bilayer component of Code E is within 0.5143 dB to the Shannon limit. At a block length of 100,000, the lower-rate code of Code E achieves a gap of 0.8 dB to the Shannon limit at BER $= 10^{-4}$. Code F has a convergence threshold within 0.1641 dB of the Shannon limit of the lower rate channel. At BER $= 10^{-4}$, the SNR gap of Code F is 0.6 dB to the Shannon limit. For this rate pair, the lengthening structure outperforms the expurgating structure. (The higher-rate component of Code E is fixed as a conventional LDPC code designed using the scheme described in Section III-B. The lower-rate component of Code F is fixed with a conventional LDPC code found in [41].)

It should be noted that only the BER performance of the bilayer structures are presented. The BER curves for the lower graph codes are omitted, as the lower graphs in both structures are fixed with conventional LDPC codes.

## VII. MULTILAYER LDPC CODES FOR RELAY NETWORKS

Thus far, we have focused on the single-relay channel and show that bilayer LDPC codes can be designed to approach the best decode-and-forward rate in this classical setting. In a more general setting, bilayer codes (or multilayer codes) can also be adopted for multiple-relay networks. This is the subject of this section.

Multiple-relay networks can have many different topologies. One way to generalize the decode-and-forward rate to multiple-relay networks is to impose a linear ordering on the intermediate relays, and let each relay completely decode the source's message with the help of relays prior to itself, then participate

TABLE II

BILAYER-LENGTHENED LDPC CODE FOR $\text{SNR}_- = -3.0728$ DB AND $\text{SNR}_+ = 5.6148$ DB

| Bilayer-lengthened LDPC code (B) | | |
|---|---|---|
| Degree | $\lambda_i^1$ | $\lambda_i^2$ |
| $i=2$ | 0.3227 | 0.1655 |
| $i=3$ | 0.2107 | 0.2617 |
| $i=5$ | 0 | 0.1505 |
| $i=6$ | 0.1247 | 0 |
| $i=7$ | 0.1194 | 0 |
| $i=10$ | 0 | 0.2977 |
| $i=11$ | 0 | 0.1246 |
| $i=20$ | 0.2225 | 0 |
| | $d_c = 5$ | $d_c' = 33$ |
| $R_- = 0.2871$, Threshold gap =0.2369 dB | | |
| $R_+ = 0.8932$, Threshold gap =0.1357 dB | | |

TABLE III

BILAYER-EXPURGATED LDPC CODE FOR $\text{SNR}_+ = -1.4237$ DB AND $\text{SNR}_- = -3.2972$ DB

| Bilayer-Expurgated LDPC code (C) | | |
|---|---|---|
| | $\lambda_{i,0}$ | $\lambda_{i,1}$ |
| $i=2$ | 0.2417 | 0.0496 |
| $i=3$ | 0.1702 | 0.0501 |
| $i=6$ | 0.1182 | 0 |
| $i=7$ | 0.0056 | 0.1348 |
| $i=18$ | 0 | 0.1573 |
| $i=19$ | 0.0267 | 0.0458 |
| $\eta = 0.9435$, $d_c = 6$, $d_c' = 2$ | | |
| $R_- = 0.2753$, Threshold gap=0.4612 dB | | |
| $R_+ = 0.3856$, Threshold gap=0.2162 dB | | |

TABLE IV

BILAYER-LENGTHENED LDPC CODE FOR $\text{SNR}_- = -3.0728$ DB AND $\text{SNR}_+ = -1.4438$ DB

| Bilayer-lengthened LDPC code (D) | | |
|---|---|---|
| Degree | $\lambda_i^1$ | $\lambda_i^2$ |
| $i=2$ | 0.3227 | 0.1580 |
| $i=3$ | 0.2107 | 0.2045 |
| $i=5$ | 0 | 0.0461 |
| $i=6$ | 0.1247 | 0.2171 |
| $i=7$ | 0.1194 | 0 |
| $i=12$ | 0 | 0.0058 |
| $i=13$ | 0 | 0.3685 |
| $i=20$ | 0.2225 | 0 |
| | $d_c = 5$ | $d_c' = 1$ |
| $R_- = 0.2871$, Threshold gap =0.2369 dB | | |
| $R_+ = 0.3843$, Threshold gap =0.2364 dB | | |

TABLE V

BILAYER-EXPURGATED LDPC CODE FOR $SNR_+ = 2.7330$ DB AND $SNR_- = -0.3273$ DB

| Bilayer-Expurgated LDPC code (E) | | | | | |
|---|---|---|---|---|---|
|  | $\lambda_{i,0}$ | $\lambda_{i,1}$ | $\lambda_{i,3}$ | $\lambda_{i,4}$ | $\lambda_{i,6}$ |
| $i=2$ | 0.0998 | 0.0805 | 0 | 0 | 0 |
| $i=3$ | 0.0827 | 0.1331 | 0 | 0 | 0 |
| $i=6$ | 0 | 0.0086 | 0.0920 | 0 | 0 |
| $i=7$ | 0 | 0 | 0.1725 | 0 | 0 |
| $i=20$ | 0 | 0 | 0 | 0.0845 | 0.2463 |
| $\eta = 0.8253, d_c = 15, d'_c = 4$ | | | | | |
| $R_- = 0.4618$, Threshold gap=0.5143 dB | | | | | |
| $R_+ = 0.7000$, Threshold gap=0.08474 dB | | | | | |

TABLE VI

BILAYER-LENGTHENED LDPC CODE FOR $SNR_- = 0.0229$ DB AND $SNR_+ = 2.6122$ DB

| Bilayer-lengthened LDPC code (F) | | |
|---|---|---|
| Degree | $\lambda_i^1$ | $\lambda_i^2$ |
| $i=2$ | 0.2421 | 0.1468 |
| $i=3$ | 0.2039 | 0.2331 |
| $i=6$ | 0.1677 | 0 |
| $i=7$ | 0.0829 | 0.3039 |
| $i=8$ | 0 | 0.0298 |
| $i=19$ | 0 | 0.2864 |
| $i=20$ | 0.3034 | 0 |
|  | $d_c = 8$ | $d'_c = 6$ |
| $R_- = 0.4877$, Threshold gap=0.1641 dB | | |
| $R_+ = 0.6906$, threshold gap=0.1208 dB | | |

in transmission of the source message to subsequent relays and to the destination. The capacity of this decode-and-forward strategy has been studied in [6] and [4]. However, this is not the only possibility. In [13], the authors cast the multiple-relay network within a parity-forwarding framework, and have been able to enlarge the decode-and-forward rate of [6] and [4]. This section focuses on two-relay networks and illustrate two fundamental ways that multiple relays can help each other and help the ultimate decoding of information at the destination. The main purpose of this section is to show that practical bilayer codes can be readily applied in these cases.

## A. Cascade Bilayer Codes for Two-Relay Networks

Consider a two-relay network depicted in Fig. 12 In this case, the first relay decodes the message from the source $w_i$, then sends out a parity $s_i^1$, just as in the single-relay case. However, suppose that the channel from the source to the second relay is weak. So, the second relay is not able to decode the source's message (even with the help of $s_i^1$), although it is able to decode $s_i^1$ itself. However, for this channel, the second relay may still help the ultimate decoding at the destination by sending out parities of parities, denoted here as $s_i^2$, to help the destination decode $s_i^1$. This "helping-the-helper" strategy can be shown to be capacity-achieving for a doubly degraded network [13], and it enlarges the achievable rates in [6] and [4].

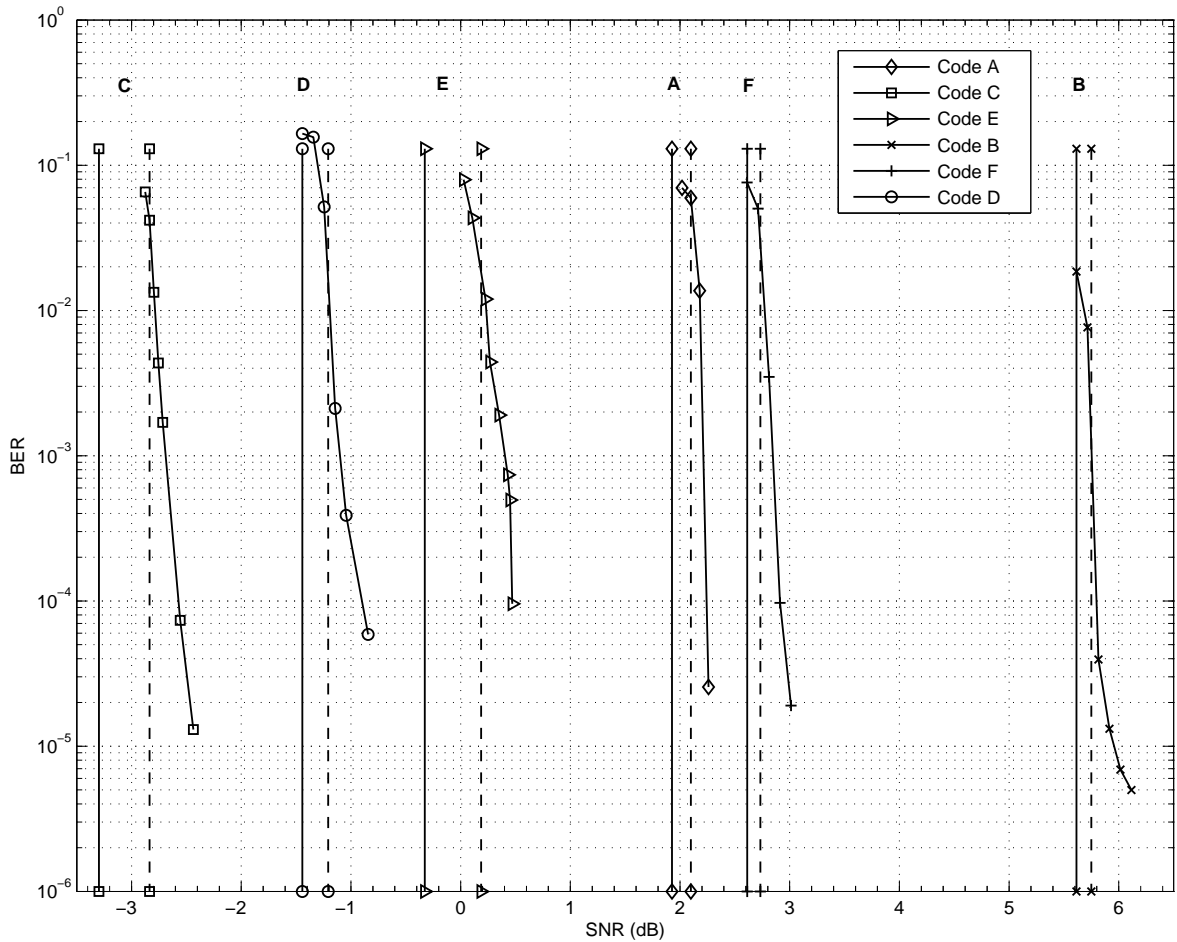

Fig. 11. Empirical bit error probability curves for the designed codes. Solid straight lines represent Shannon limits for each code, and dashed lines represent the convergence threshold computed by density evolution.

The code construction for this relay network is shown in Fig. 13. It consists of a cascade of two bilayer codes. The source message is coded by a bilayer code $\mathcal{C}_1$. Upon decoding $\mathcal{C}_1$, the first relay computes additional parities for $\mathcal{C}_1$ and re-encodes them using $\mathcal{C}_2$, which is another bilayer code. The second relay decodes $\mathcal{C}_2$, then computes extra parities for $\mathcal{C}_2$ and re-encodes them using $\mathcal{C}_3$. Finally, the destination first decodes $\mathcal{C}_3$ to recover the extra parities needed to decode $\mathcal{C}_2$. Then, it decodes $\mathcal{C}_2$ to recover the parities

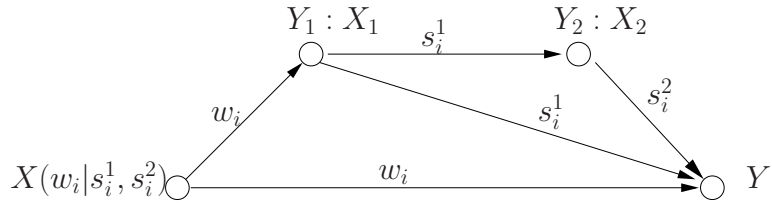

Fig. 12. A two-relay network in which the second relay facilitates the transmission of parity bits from the first relay to the destination

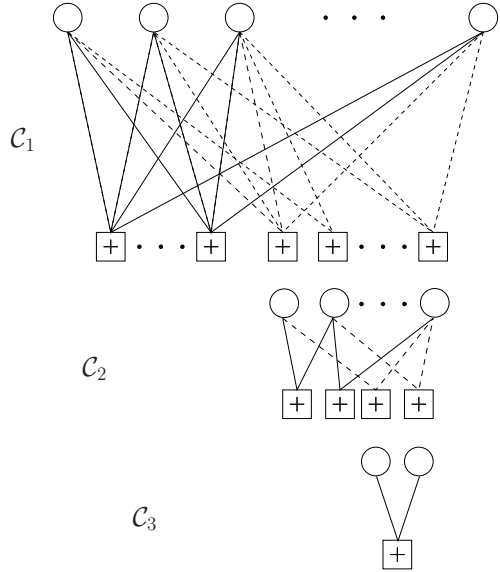

Fig. 13. Cascaded bilayer codes for the two-relay network in Fig. 12

of $\mathcal{C}_1$. Finally, the destination decodes $\mathcal{C}_1$.

Clearly, the bilayer codes that have been devised for single-relay channels can be directly cascaded to design coding systems capable of approaching the best achievable rate in this network.

## B. Doubly Bilayer Codes for Two-Relay Networks

Consider a different two-layer network depicted in Fig. 14 in which the channel between the first relay and the destination is weak. In this case, the optimal strategy is for the first relay to help the second relay, so that the second relay can ultimately help the destination.

The code construction for this relay network is shown in Fig. 15. It is a doubly bilayer code in the following sense. The source encodes its message using $\mathcal{C}_1$. The first relay decodes $w_i$, computes $k_2$ parities bits, and re-encode the parities using $\mathcal{C}_2$ for the second relay. The second relay decodes $w_i$ with the help of $k_2$ parities. Then, it computes separate $k_3$ parity bits to be re-encoded by $\mathcal{C}_3$. The destination decodes $\mathcal{C}_3$, then $\mathcal{C}_1$, the source message. The achievable rates using the above strategy is a special case of the achievable rate in [6].

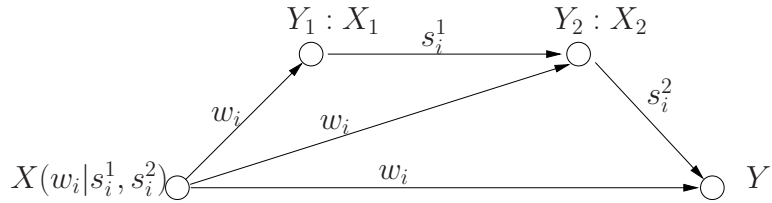

Fig. 14. A two-relay network in which the first relay helps the second relay to decode the source message

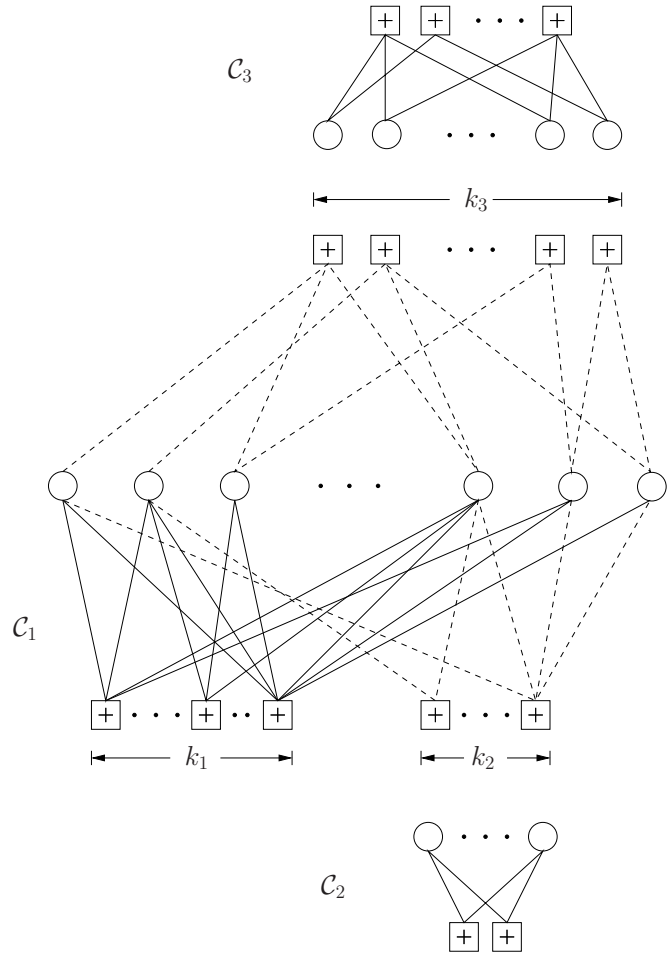

Fig. 15. Doubly bilayer codes for the two-relay network in Fig. 14

For this relay network, $\mathcal{C}_2$ and $\mathcal{C}_3$ are conventional LDPC codes. However, $\mathcal{C}_1$ must be specially designed as two bilayer codes extended from the same base code. The code design methodology described in the previous section can again be used for this network. For example, Code A and Code E can be utilized to construct codebooks for implementing this protocol with a source rate of $R = 0.7$. The first relay decodes the source codeword at $R = 0.7$; the second relay, with the help of $k_2$ parity bits from the first relay at a rate $0.7 - 0.6363$, can use Code A to decode the source codeword. The second relay then sends out $k_3$ parity bits to the destination at a rate $0.7 - 0.4618$, which enables the destination to decode the source

383838

codeword using Code E.

## VIII. Concluding Remarks

Binning is of fundamental importance in multiuser information theory. This paper provides a practical implementation of the binning strategy for the relay channel from a linear coding perspective, in which extra information is generated at the relay to facilitate the overall communication between the source and the destination. A key feature of the code design is the construction of a bilayer LDPC code that is capable of approaching the Gaussian channel capacity at two different SNRs and at two different rates. We show that conventional code design techniques must be significantly modified for the design of these multirate codes in order to achieve capacity-approaching performances.

The code construction in this paper shows that the binning operation for the relay channel is fundamentally easier to implement in practice than the binning techniques for sources and channels with side information. The former is an error-correcting problem; the latter essentially a quantization problem for which efficient coding methods are not yet known.

The concept of bilayer codes can be extended to relay networks in which cascades of bilayer codes and multilayers LDPC code may be needed. While in principle these codes can be designed and optimized for a given network topology, as the network size grows, the encoding and decoding protocols become increasingly complex, and the tuning of the code parameters increasingly involved. The code structure illustrated in this paper suggests that practical protocols for the relay network should involve universal and rateless codes. The bilayer code design methodology described in this paper is a first step toward this goal.

## Acknowledgment

The authors would like to thank Masoud Ardakani, for insightful discussions on LDPC code optimization and for providing his softwares.